# An energy landscape approach to locomotor transitions in complex 3-D terrain

Ratan Othayoth, George Thoms, Chen Li*

Department of Mechanical Engineering, Johns Hopkins University

3400 N. Charles Street, Baltimore, Maryland 21218, USA

*Corresponding author. Email: [redacted]

**Author Contributions**

R.O. designed study, developed robotic physical model, performed animal and robot experiments, analyzed data, developed energy landscape model, drafted and revised the paper; G.T. developed robotic physical model and performed preliminary robot experiments; C.L. designed and oversaw study, defined analyses, and wrote and revised the paper.

**This PDF file includes:**

## Abstract

Effective locomotion in nature happens by transitioning across multiple modes (e.g., walk, run, climb). Despite this, far more mechanistic understanding of terrestrial locomotion has been on how to generate and stabilize around near-steady-state movement in a single mode. We still know little about how locomotor transitions emerge from physical interaction with complex terrain. Consequently, robots largely rely on geometric maps to avoid obstacles, not traverse them. Recent studies revealed that locomotor transitions in complex 3-D terrain occur probabilistically via multiple pathways. Here, we show that an energy landscape approach elucidates the underlying physical principles. We discovered that locomotor transitions of animals and robots self-propelled through complex 3-D terrain correspond to barrier-crossing transitions on a potential energy landscape. Locomotor modes are attracted to landscape basins separated by potential energy barriers. Kinetic energy fluctuation from oscillatory self-propulsion helps the system stochastically escape from one basin and reach another to make transitions. Escape is more likely towards lower barrier direction. These principles are surprisingly similar to those of near-equilibrium, microscopic systems. Analogous to free energy landscapes for multi-pathway protein folding transitions, our energy landscape approach from first principles is the beginning of a statistical physics theory of multi-pathway locomotor transitions in complex terrain. This will not only help understand how the organization of animal behavior emerges from multi-scale interactions between their neural and mechanical systems and the physical environment, but also guide robot design, control, and planning over the large, intractable locomotor-terrain parameter space to generate robust locomotor transitions through the real world.

## Significance statement

Effective locomotion in nature happens by transitioning across multiple modes (e.g., walk, run, climb). Using lab experiments on a model system, we demonstrate that an energy landscape approach helps understand how multi-pathway transitions across locomotor modes in complex 3-D terrain statistically emerge from physical interaction. Animals' and robots' locomotor modes are attracted to basins of a potential energy landscape. They can use kinetic energy fluctuation from oscillatory self-propulsion to cross potential energy barriers, escaping from one basin and reaching another to make locomotor transitions. Our first-principle energy landscape approach is the beginning of a statistical physics theory of locomotor transitions in complex terrain. It will help understand and predict how animals, and how robots should, move through the real world.

## Main Text

To move about in the environment, animals can use many modes[*] of locomotion (e.g., walk, run, crawl, climb, fly, swim, jump, burrow) (1–3) and must often transition across them (4, 5) (e.g., Fig. 1*A*, SI Appendix, Movie S1). Despite this, far more of our mechanistic understanding of terrestrial locomotion has been on how animals generate (6–10) and stabilize (11–14) steady-state, limit-cycle-like locomotion using a single mode.

Recent studies begin to reveal how terrestrial animals transition across locomotor modes in complex environments. Locomotor transitions, like other animal behavior, emerge from multi-scale interactions of the animal and external environment across the neural, postural, navigational, and ecological levels (15–17). At the neural level, terrestrial animals can use central pattern generators (18) and sensory information (19–21) to switch locomotor modes to traverse different media or overcome obstacles. At the ecological level, terrestrial animals foraging across natural landscapes switch locomotor modes to minimize metabolic cost (22). At the intermediate level, terrestrial animals also transition between walking and running to save energy (23). However, there remains a knowledge gap in how locomotor transitions in complex terrain emerge from direct physical interaction (i.e., terradynamics (24)) of an animal's body and appendages with the environment. In particular, we lack theoretical concepts for thinking about how to generate and control locomotor transitions in complex terrain that are on the same level of limit cycles for single-mode locomotion (25). For example, locomotion in irregular terrain with repeated perturbations is rarely near steady state and requires an animal to continually modify its behavior, which cannot be well described by limit cycles (26, 27).

[*] Here we use mode in the general sense to refer to distinct, stereotyped locomotor behavior, not confined to limit-cycle behavior such as gaits or templates (25).

Understanding of how to make use of physical interaction with complex terrain (environmental affordance (28, 29)) to generate and control locomotor transitions is also critical to advancing mobile robotics. Similar to personal computers decades ago, mobile robots are on the verge of becoming a part of society. Some robots (e.g., robot vacuums, self-driving cars) already excel at navigating flat surfaces, by transitioning across driving modes (e.g., forward drive, U-turn, stop, park (30)) to avoid sparse obstacles using a geometric map of the environment (31). However, many critical applications, such as search and rescue in rubble, inspection and monitoring in buildings, extraterrestrial exploration through rocks, and even drug delivery inside a human body, require robots to transition across diverse locomotor modes to traverse unavoidable obstacles in complex terrain (4, 5, 32) (Fig. 1*B*). Yet, terrestrial robots still struggle to do so robustly (33), because we do not understand well how locomotor transitions (or lack thereof) emerge from physical interaction with complex terrain.

Our study is motivated by recent observations in a model system of insects traversing complex 3-D terrain. The discoid cockroach, native to rainforest floor, can traverse flexible, grass-like beam obstacles using many locomotor modes, stochastically transitioning across them via multiple pathways (34). For simplicity, hereafter we focus on the transition between two modes. The animal often first pushes against the beams, and beam elastic restoring forces lead the animal body to pitch up (Fig. 1*C*, blue). After this, though, the animal rarely pushes across (3% probability) but often rolls (Fig. 1*C*, red) to maneuver through beam gaps (45% probability). We define these as "pitch" and "roll" modes. Note that we use "locomotor mode" here in the general sense, not confined to limit-cycle locomotor behavior. The pitch mode is more challenging than the roll mode because the animal has to lift its weight and deflect the beams more (this is only true when beams are stiff, though; see Results). Thus, the animal appears to statistically transition from less to more favorable modes. In addition, the animal's body oscillates as its legs continually pushed against the ground when trying to traverse. Besides in obstacle traversal, similar multi-pathway locomotor transitions, preference of some modes over others, and seemingly wasteful body oscillation were observed in self-righting of insects (35).

In the field of protein folding, adopting a statistical physics view and using an energy landscape approach led researchers to recognize that proteins fold via multiple pathways and understand the physical principles (36–38). These near-equilibrium, microscopic systems statistically transition from higher to lower energy states (local minima) on a free energy landscape (increasing thermodynamic favorability). Thermal fluctuation helps the system stochastically cross energy barriers at transition states (saddle points between local minimum basins). These physical principles operating on a rugged landscape leads to the multi-pathway protein folding transitions. Inspired by the seeming similarities of our system to them, we contend that an energy landscape approach helps understand how self-propelled, far-from-equilibrium macroscopic animals' and robots' probabilistic locomotor transitions in complex 3-D terrain emerge from physical interaction, whose equations of motion are unknown or intractable (39, 40). Specifically, we hypothesize that:

(1) The self-propelled system's state is attracted to a local minimum basin on a potential energy landscape; locomotor transition from one mode to another can be viewed as the system state escaping from one basin and settling into another. (What governs transition?)

(2) When it is comparable to the potential barrier, kinetic energy fluctuation from oscillatory self-propulsion helps the system escape from a landscape basin to make locomotor transitions. (When does transition happen?)

(3) Escape from a basin is more likely towards a direction along which the escape barrier is lower. (How does transition happen?)

To begin to establish an energy landscape approach of locomotor transitions across modes in complex 3-D terrain, we tested these hypotheses for the two representative modes (pitch and roll) of the model body-beam interaction system defined above. Although the previous study introduced an early energy landscape model to qualitatively explain why locomotor shape affected physical interaction and thus locomotion (34), none of these hypotheses were proposed or tested. We emphasize that our potential energy landscape directly arises from locomotor-terrain interaction physics using first principles. This is unlike artificially defined potential functions to explain walk-to-run transition (41) and other non-equilibrium biological phase transitions (42), or metabolic energy landscapes inferred from oxygen consumption measurements to explain behavioral switching of locomotor modes (22).

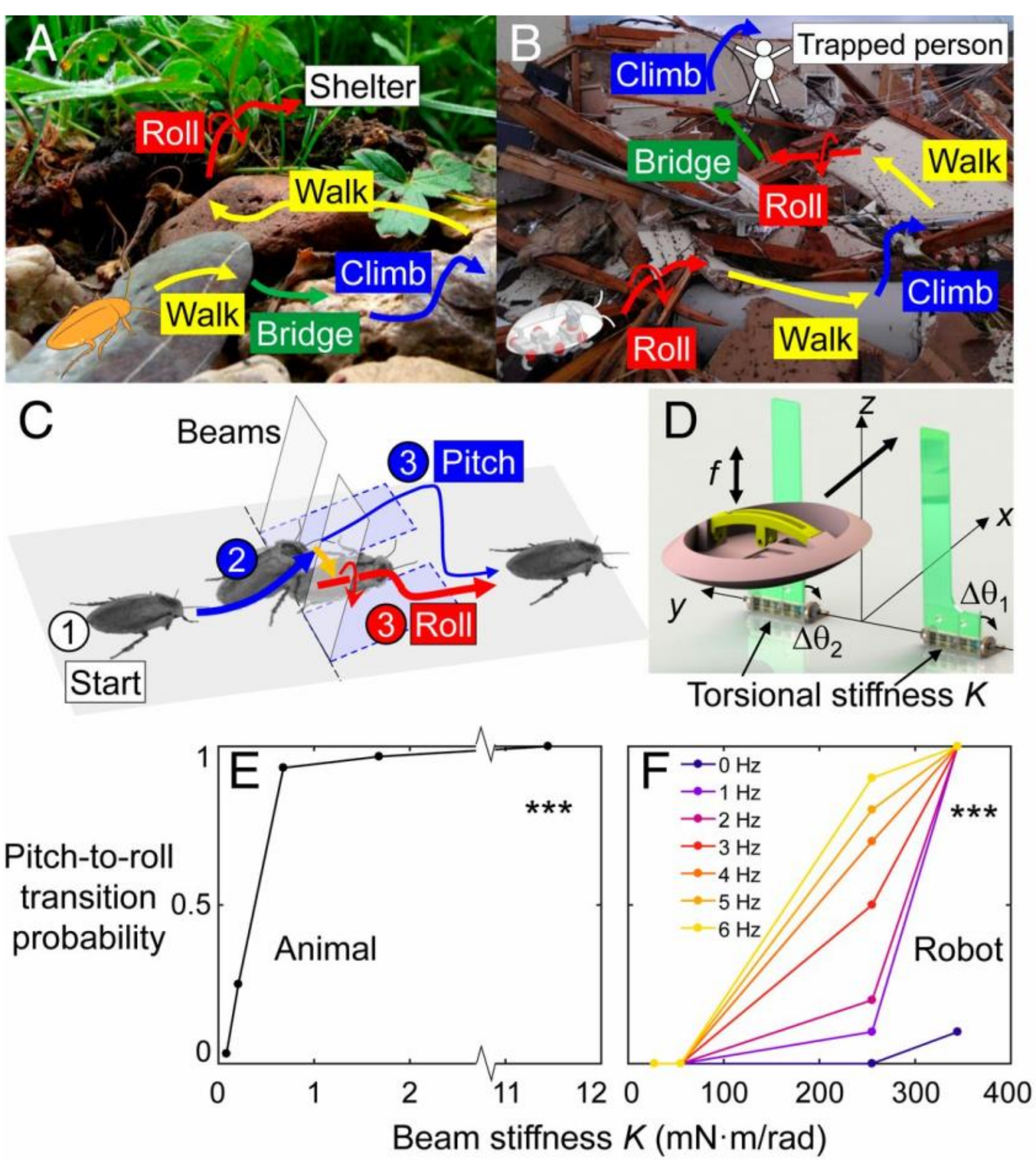

**Fig. 1.** Locomotor transitions of animals and robots in complex terrain. (*A*, *B*) Illustrative locomotor transitions of (*A*) a cockroach traversing forest floor (photo: Scott Brill, seekraz.wordpress.com, with permission) and (*B*) a robot traversing rubble for search and rescue. (*C*) A cockroach transitioning

(orange arrow) from pitch to roll mode to traverse grass-like beam obstacles. (*D*) Robotic physical model. (*E*, *F*) Pitch-to-roll transition probability of animal (*E*) and robot (*F*) as a function of beam stiffness $K^{\dagger}$. For robot, we varied oscillation frequency $f$ to vary kinetic energy fluctuation. *** indicates a significant dependence on $K$ (animal: mixed-effects chi-squared test, $P < 0.0001$, $\chi^2 = 297.4$; robot: chi-squared test, $P < 0.0001$, $\chi^2 = 247.1$). $n$ = 64, 60, 60, 62, 64 trials for animal and $n$ = 70 trials at each $K$ for robot.

Because animal locomotion emerges from complex interactions of neural and physical mechanisms (1), to observe the outcome of pure physical interaction, we developed and tested a minimalistic robotic physical model (Fig. 1*D*, SI Appendix, Movie S2) with feedforward control. The robot had an ellipsoid-like body that was propelled forward at a constant speed and was free to pitch and roll (achieved through a gyroscope mechanism) in response to interaction with two beams. The body was constrained not to yaw or move laterally to simplify energy landscape modeling. We also performed experiments with the discoid cockroach traversing beams during escape response to study how physical interaction affects the animal's locomotor transitions when neural control is bandwidth limited (1). Comparison of robot and animal observations can reveal aspects of the transitions that likely involve neural mechanisms.

To test the first hypothesis, in both robot and animal experiments, we used rigid "beams" with torsional joints at the base (SI Appendix, Figs. S1, S2) as one-degree-of-freedom 3-D terrain components to generate a simple potential energy landscape. We then reconstructed the potential energy landscape and 3-D motion of the robot or animal body and beams in high accuracy (as opposed to visual examination in the previous study (34)) (SI Appendix, Figs. S3, S4, Movies S3, S4) for the entire traversal. This allowed us to quantify how the system state behaved on the landscape during each observed locomotor mode and transition between modes. To test the second hypothesis, for the robot, we applied controlled oscillation with variable frequency $f$ to vary kinetic energy fluctuation (SI Appendix, Fig. S5). Because we could not vary the animal's naturally occurring body oscillation, in animal experiments we changed the barrier relative to kinetic energy oscillation by varying beam torsional joint stiffness $K$ by over an order of magnitude in the range of natural flexible terrain elements (SI Appendix, Table S2). $K$ was also varied by over an order of magnitude for robot experiments and, together with animal experiments, helped elucidate how transition depended on terrain properties. Because the potential energy landscape consists of not only beam elastic energy but also body and beam gravitational energy, variation of $K$ also changed how escape barrier compared in different directions, allowing the third hypothesis to be tested. See Methods and SI Appendix, Supplementary Methods for technical detail and SI Appendix, Table S1 for sample sizes.

---

[†]At $f$ = 0, transition occurred in one trial at $K$ = 344 mN·m/rad (resulting in a 10% probability) due to lateral displacement of the body. This was from lateral bending of the vertical bar driving the body forward due to large lateral force from the stiff beams, an effect not captured by our model.

**Results**

Before encountering the beams, both the robot and animal moved forward with a near horizontal body posture. After beam contact, both the robot and animal started traversing by pushing against the beams, with the body pitched up. As beam stiffness $K$ increased, pitch-to-roll transition probability increased for both the robot and animal (Fig. 1*E*, *F*; $P < 0.0001$, mixed-design chi-squared test). At low $K$, neither transitioned to the roll mode even with body oscillation. At the highest $K$, both always transitioned, except for the robot without oscillation. In addition, for the robot at high $K$ (255 mN·m/rad), pitch-to-roll transition probability increased with oscillation frequency $f$ (Fig. 1*F*) and thus with kinetic energy fluctuation (SI Appendix, Fig. S5*A*). At the highest $K$ tested (344 mN·m/rad), pitch-to-roll transition probability reached one for all $f > 0$ tested. For simplicity, below we first describe robot results followed by animal results.

We tested the first hypothesis by reconstructing the robot's potential energy landscape and evaluating how its system state behaved on the landscape (Fig. 2, SI Appendix, Movie S4). Using the measured physical and geometric parameters of the body and beams, we calculated the robot's system potential energy (sum of body and beam gravitational energy and beam elastic energy) as a function of body pitch, roll, and forward position $x$ relative to the beams. For simplicity, we first examine results at $K$

= 255 N·m/rad. Before the body contacted the beams (Fig. 2*A*, i), pitching or rolling increased body gravitational energy (because body center of mass was below rotation axes, SI Appendix, Fig. S6). Thus, the potential energy landscape over body pitch-roll space had a global minimum at zero pitch and zero roll, i.e., when the body was horizontal (Fig. 2*B*, i). As the body moved closer and interacted with the beams (Fig. 2*A*, ii, iii), the global minimum evolved into a "pitch" local minimum at a finite pitch and zero roll (Fig. 2*B*, ii, iii, blue). Meanwhile, two "roll" local minima emerged at near zero pitch and a finite positive or negative roll (Fig. 2*B*, ii, iii, red, for rolling right or left), whose energies were lower than the pitch local minimum. Hereafter, we refer to these local minimum basins as pitch and roll basins[‡].

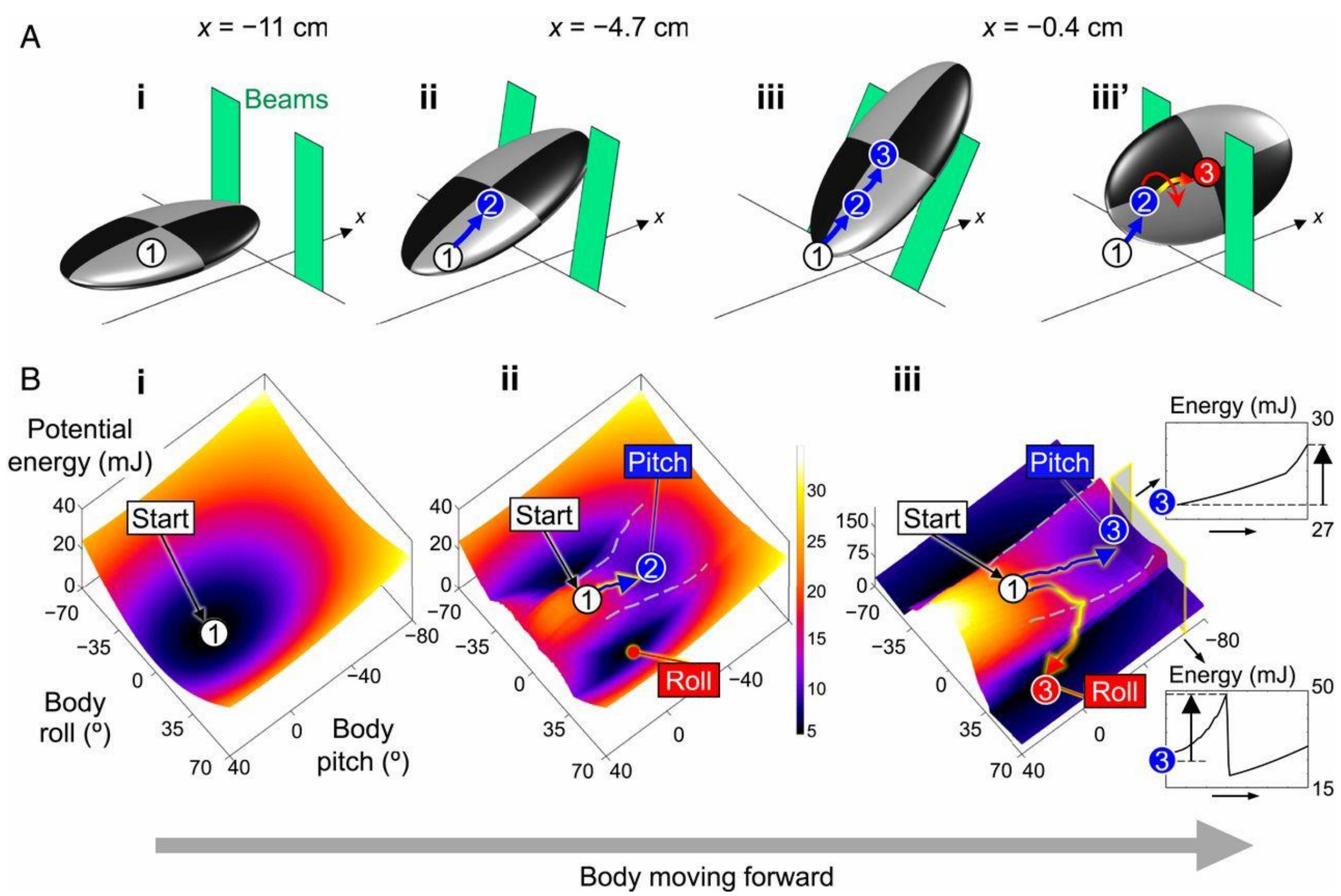

**Fig. 2.** Robot locomotor transitions on a potential energy landscape. Results are shown at $K$ = 255 mN·m/rad. (*A*) Snapshots of body before and during interaction with two beams in pitch (i, ii, iii) and

‡A fourth basin also emerged with its local minimum at a finite positive pitch and zero roll, corresponding to the body pitching down against the beams. However, such a configuration was never observed in the robot or animal.

roll (iii') modes. (*B*) Snapshots of landscape over body pitch-roll space before (i) and during (ii, iii) interaction. Representative system state trajectories are shown for being trapped in pitch basin (blue) and transitioning to roll basin (red). Insets in (iii) define potential energy barriers to escape from pitch local minimum in pitch-up and positive roll directions. Dashed gray curves on landscape show boundaries between pitch and roll basins. Note that landscape evolves as body moves forward (increasing $x$), and only part of landscape over pitch-roll space is shown to focus on pitch and roll basins.

We discovered that the robot's system state during the observed pitch and roll modes were attracted to the pitch and roll basins, respectively. When the body was far away from the beams, the system state in pitch and roll space settled to the global minimum of the landscape (Fig. 2*B*, i, SI Appendix, Movie S4). During beam interaction, without oscillation, the system state was trapped in the pitch basin, leading to the body pushing across the beams in a pitched-up orientation with little roll (Figs. 2*A*, *B*, ii, iii, SI Appendix, Movie S4, top). With oscillation, the system stochastically escaped from the pitch basin and crossed a potential energy barrier to reach the roll basin (Fig. 2*B*, iii, SI Appendix, Movie S4, bottom), thereby transitioning from the pitch to the roll mode (Fig. 2*B*, ii, iii'). We examined system state trajectory on the landscape reconstructed for each trial (see examples in SI Appendix, Movie S6, third row). Whether the robot was trapped in the pitch mode (blue trajectories) or transitioned to the roll mode (red trajectories), its system state was attracted to the corresponding basin in nearly all trials (99%, not significantly different from 1, $P > 0.15$, Student's $t$-test, Fig. 4*A*, iii). Because of this strong attraction, the measured system potential energy closely matched the observed mode basin's local minimum energy throughout traversal (Fig. 5iii, solid vs. dashed curves). All these findings held true at other $K$ (near 100%, Figs. 4*A*, Fig. 5, SI Appendix, Movie S6). Together, these robot results supported our first hypothesis.

Next, we tested the second hypothesis. We first observed how kinetic energy fluctuation affected the robot's escape from a basin. Again, we examine results at $K$ = 255 N·m/rad first for simplicity. As $f$ increased (which increased kinetic energy fluctuation), the system was more likely to escape from the pitch basin it was initially attracted to and reach the roll basin (Fig. 3, SI Appendix, Movie S5), resulting in more likely pitch-to-roll transitions (Fig. 1*F*, *K* = 255 mN·m/rad). Then, we compared the minimal potential energy barrier to escape from the pitch local minimum with the average kinetic energy fluctuation at *f* = 6 Hz (Fig. 4*C*, iii, SI Appendix, Movie S7, bottom). The escape barrier depended on both towards which direction the system moved in the pitch-roll space (Fig. 2*B*, iii, insets, Fig. 4*B*, iii) and body forward position *x* relative to the beams (Fig. 4*C*, iii, SI Appendix, Movie S7, bottom). Minimal escape barrier occurred at the saddle point between the pitch and roll basins (Fig. 4*C*, yellow dot), which we defined as pitch-to-roll transition barrier. Only within a small range of *x* was average kinetic energy fluctuation at *f* = 6 Hz (Fig. 4*C*, iii, green) sufficient for overcoming pitch-to-roll transition barrier (Fig. 4*C*, iii, black, SI Appendix, Movie S8, third column). This range matched remarkably well with the *x* range over which pitch-to-roll transition was observed with increasing likelihood with *f* (gray band showing mean ± s.d. from all trials across *f*). All these findings held true at *K* = 344 N·m/rad. At *K* = 28 N·m/rad, minimal escape barrier far exceeded kinetic energy fluctuation, consistent with the absence of transition. Together, these robot results supported our second hypothesis.

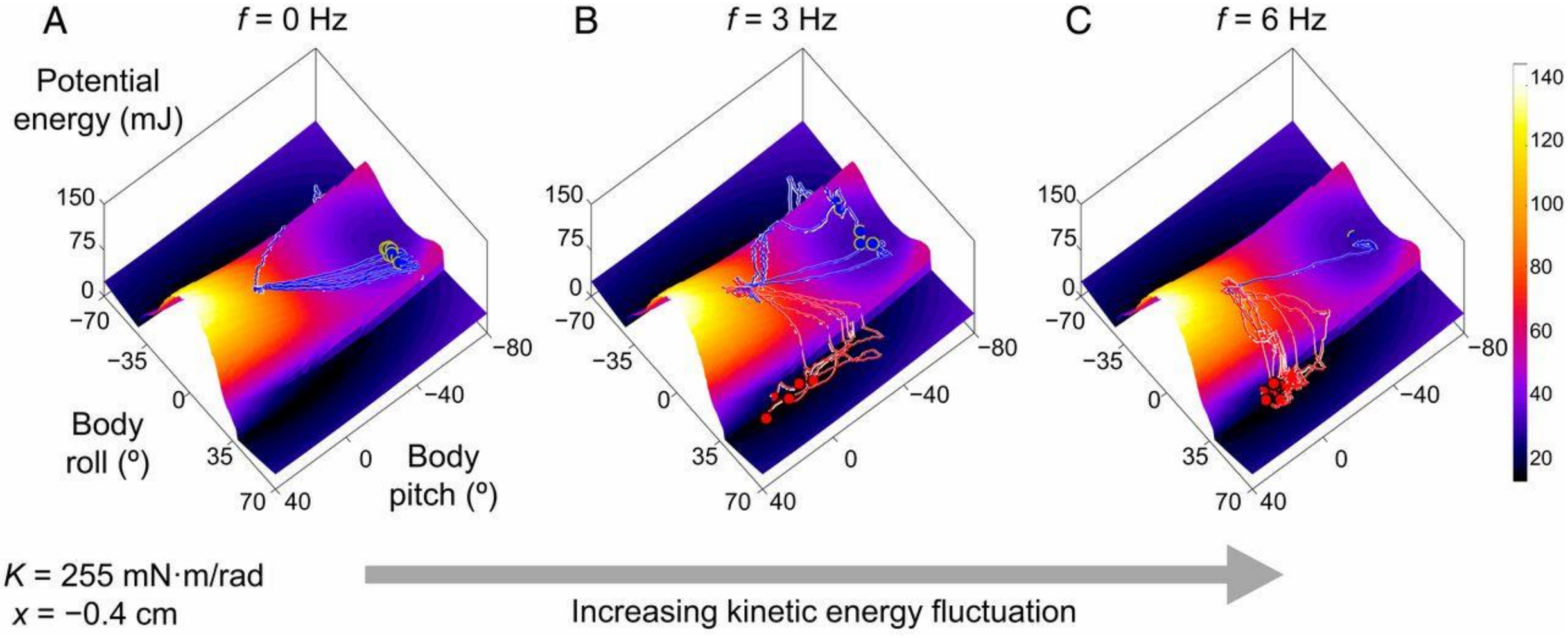

**Fig. 3.** Robot locomotor transitions are stochastic and become more likely as kinetic energy fluctuation increases. Comparison of state trajectory ensemble on average landscape (snapshot at $x = -0.4$ cm) across oscillation frequencies: (*A*) *f* = 0 Hz; (*B*) *f* = 3 Hz; (*C*) *f* = 6 Hz. Results are shown at *K* = 255 mN·m/rad. Blue and red curves show trials trapped in pitch basin and transitioning to roll basin, respectively. Trials

in which body rolls left are flipped to rolling right considering lateral symmetry. $n$ = 10 trials at each $f$. Only part of landscape over pitch-roll space is shown to focus on pitch and roll basins. Blue trajectories exiting pitch basin is an artifact of landscape averaging.

Finally, we tested the third hypothesis by examining the direction towards which the robot's system state moved during interaction. At each $K$, when the body was not in contact with the beams, the escape barrier was large along all directions in the pitch-roll space (SI Appendix, Movie S7, bottom, Movie S8, second row, e.g., $x = -80$ mm). As the body moved forward (increasing $x$), the escape barrier towards the direction of roll basins reduced drastically, becoming comparable to or even smaller than average kinetic energy fluctuation at $f$ = 6 Hz (green circle) at the saddle point (yellow dot). By contrast, escape barrier in the direction of pitching up or down was always greater than average kinetic energy fluctuation (Fig. 4*B*, SI Appendix, Movie S8, third row). Examination of how the system state moved on the landscape (SI Appendix, Movie S9, top) and probability distribution of system state velocity directions in the pitch-roll space (Fig. 4*D*, SI Appendix, Movie S9, bottom) showed that escape was more aligned with the direction of the saddle point between pitch and roll basins, i.e., escape was more likely towards the direction of lower barrier. This is intuitive because in other directions escape barrier was higher and often exceeded kinetic energy fluctuation. Together, these robot observations supported our third hypothesis.

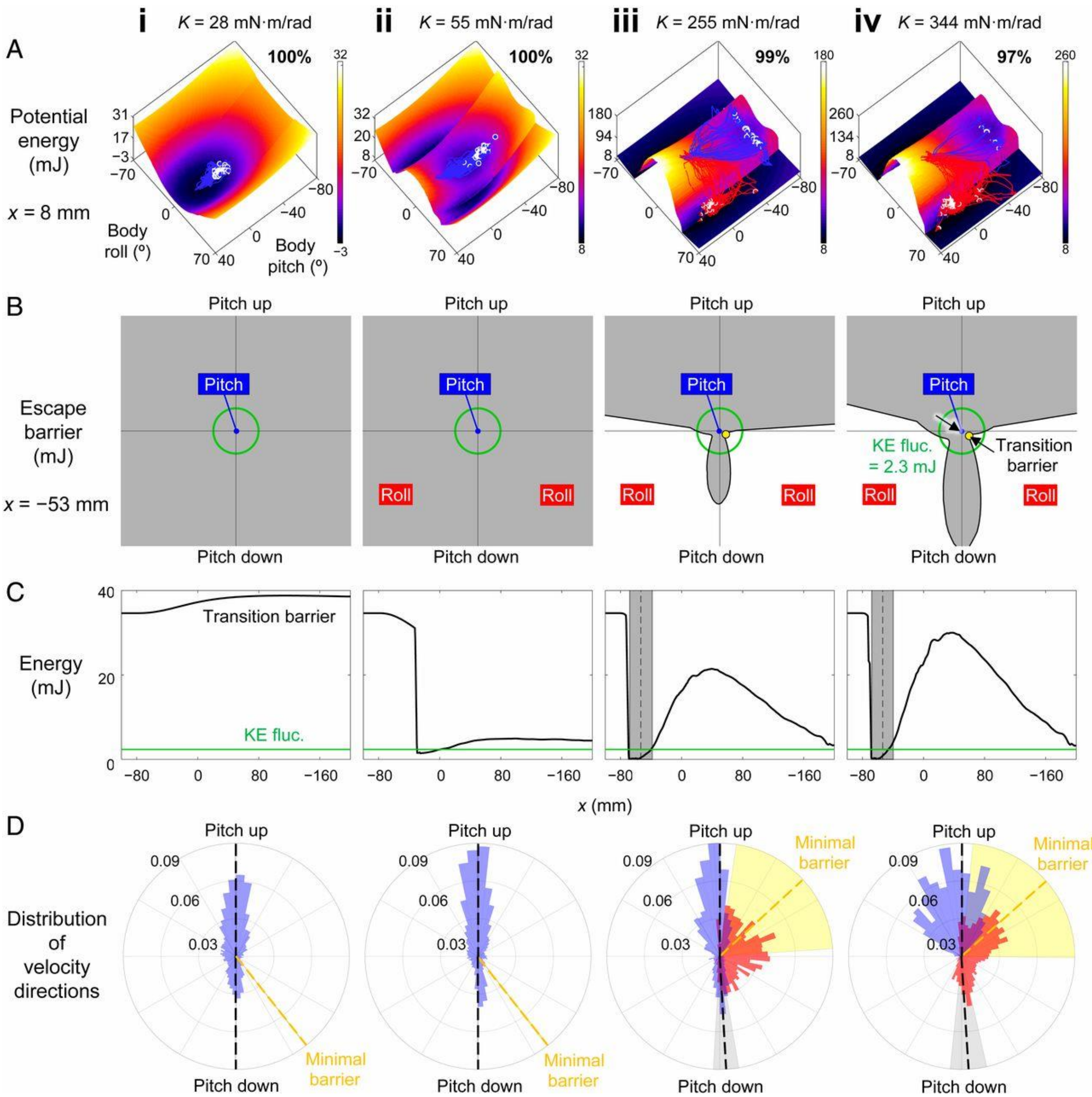

**Fig. 4.** Robot tends to transition to roll basin when kinetic energy fluctuation is comparable to potential energy barrier to escape pitch local minimum and towards direction of lower barrier. (*A*) Average potential energy landscape over pitch-roll space (snapshot at $x$ = 8 mm) with ensemble of state trajectories. Blue and red curves show trials trapped in pitch basin and transitioning to roll basin, respectively. Note that landscape evolves as body moves forward (increasing $x$) (SI Appendix, Movie S8) and only part of the landscape over pitch-roll space is shown to focus on the pitch and roll basins. Top right number on each landscape shows percentage of trials in which system state is attracted to pitch/roll basin corresponding to observed mode. Blue trajectories exiting pitch basin is an artifact of landscape averaging. (*B*) Polar plot of potential energy barrier to escape from pitch local minimum (blue dot) along all directions in pitch-roll space (snapshot at $x = -53$ mm). Pitch-to-roll transition barrier is defined as minimal escape barrier (arrows in iv), which occurs at saddle point between pitch and roll basins (yellow dot). (*C*) Pitch-to-roll transition barrier as a function of $x$. Gray band shows $x$ range in which pitch-to-roll transition is observed (mean ± s.d.). Green circle/line in *B*, *D* shows measured average kinetic energy fluctuation of 2.3 mJ at highest $f = 6$ Hz tested (SI Appendix, Fig. S5*A*). (*D*) Probability distribution of state velocity directions in pitch-roll space in the $x$ range where transition is observed (gray band in *C*). Blue and red are data from trials trapped in pitch basin and transitioning to roll basin, respectively. Trials in which body rolls left are flipped to rolling right considering lateral symmetry. Black dashed lines and gray shaded sectors show angular direction of maximal escape barriers (mean ± s.d) along pitch up and down directions. Yellow dashed line and shaded sector show angular direction of minimal escape barrier (mean ± s.d), which occurs at saddle point. Columns i-iv are at $K$ = 28, 55, 255, and 344 mN·m/rad. Data shown in *A*, *C*, and *D* are for all $f$ tested ($n$ = 70 trials) at each $K$.

Comparison of robot observations across $K$ further suggested a concept of favorability for locomotor transitions. As $K$ increased, pitch-to-roll transition became more likely (Fig. 4*A*), saturating at one for all $f > 0$ tested at the highest $K$ (Fig. 1*F*). Intuitively, when the beams were flimsy, the body pushed across (trapped in the pitch mode) as if nothing were there; when the beams were rigid, the body could not push across and must roll. Thus, the likelihood of pitch-to-roll transition is positively correlated with how favorable transitioning to the roll mode is relative to staying in the pitch mode. To provide a measure of favorability, we compared whether the pitch or roll basin was lower during traversal, measured at their respective local minimum (Fig. 5, Movie S8, fourth row). At low $K$ (28 mN·m/rad), the pitch basin remained the global minimum basin throughout traversal (Fig. 5i), indicating that the pitch mode was more favorable. As $K$ increased, the pitch basin became increasingly higher than the roll basin (Fig. 5ii-iv), indicating that the roll mode became increasingly more favorable. At small $K$ = 55 mN·m/rad for $x > 0$, although the roll mode was more favorable (Fig. 5ii), kinetic energy fluctuation was smaller than the transition barrier (Fig. 4*C*, ii); thus, transition did not occur (Fig. 4*A*, ii). We emphasize that the negative correlation between the probability of staying in or transitioning to a mode and its relative basin height is only an emergent outcome of the transition physics. The passive robot does not directly feel how high or how low an adjacent basin is; whether it escapes and makes a transition only depends on the basin in which it currently resides. Exactly how favorability difference between basins emerges from the local dynamics of escaping from a basin remains to be understood.

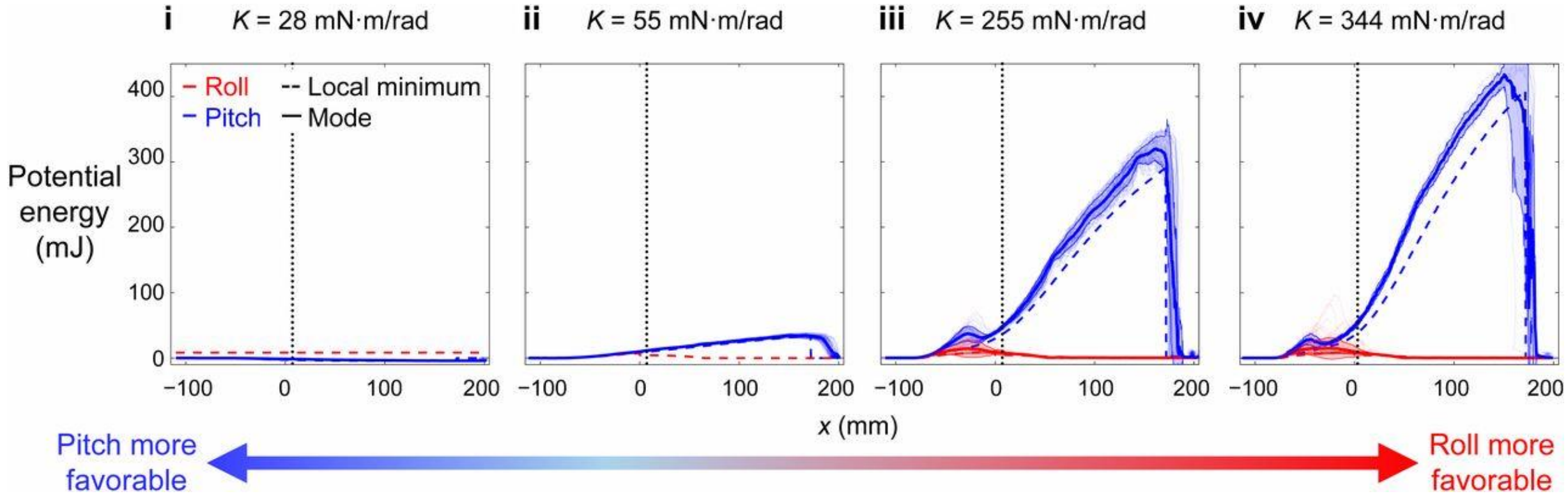

**Fig. 5.** Favorability measure for robot. Potential energy of measured pitch and roll modes (solid, mean ± s.d.) and of pitch and roll local minima[§] (dashed) as a function of $x$. Measured data are for all $f$ tested ($n$ = 70 trials) at each $K$. Blue and red show trials trapped in pitch basin and transitioning to roll basin, respectively. Columns i-iv are at $K$ = 28, 55, 255, and 344 mN·m/rad. Dotted line at $x$ = 8 mm shows location of snapshots in Fig. 4*A*.

Similar to the feedforward-controlled robot, the animal's system state during the observed pitch or roll mode was attracted to the corresponding basin of the potential energy landscape (Fig. 6*A*, ~90% of trials at all *K*; SI Appendix, Movie S10, top and middle). In addition, pitch-to-roll transition mostly occurred when *both* average kinetic energy fluctuation became comparable to transition barrier *and* the roll mode became more favorable than the pitch mode (Fig. 6*B*, SI Appendix, Movie S10, bottom). These similar observations were remarkable because, for the animal that displayed larger lateral motion and yawing, leg motion, and individual variation, the landscape (which was averaged from all trials) provided a much coarser approximation of the system than for the simpler, well-controlled robot. These animal results supported our first and second hypotheses. We did not test the third hypothesis in the animal, considering that the measured system state velocity was noisy and the animal had higher lateral and yaw motion during traversal.

[§]At $K$ = 28 mN·m/rad, roll local minimum does not exist. For comparison with other $K$, we defined it to be at (pitch, roll) = (0°, ± 42°) based on the minimal body roll required to traverse without beam deflection.

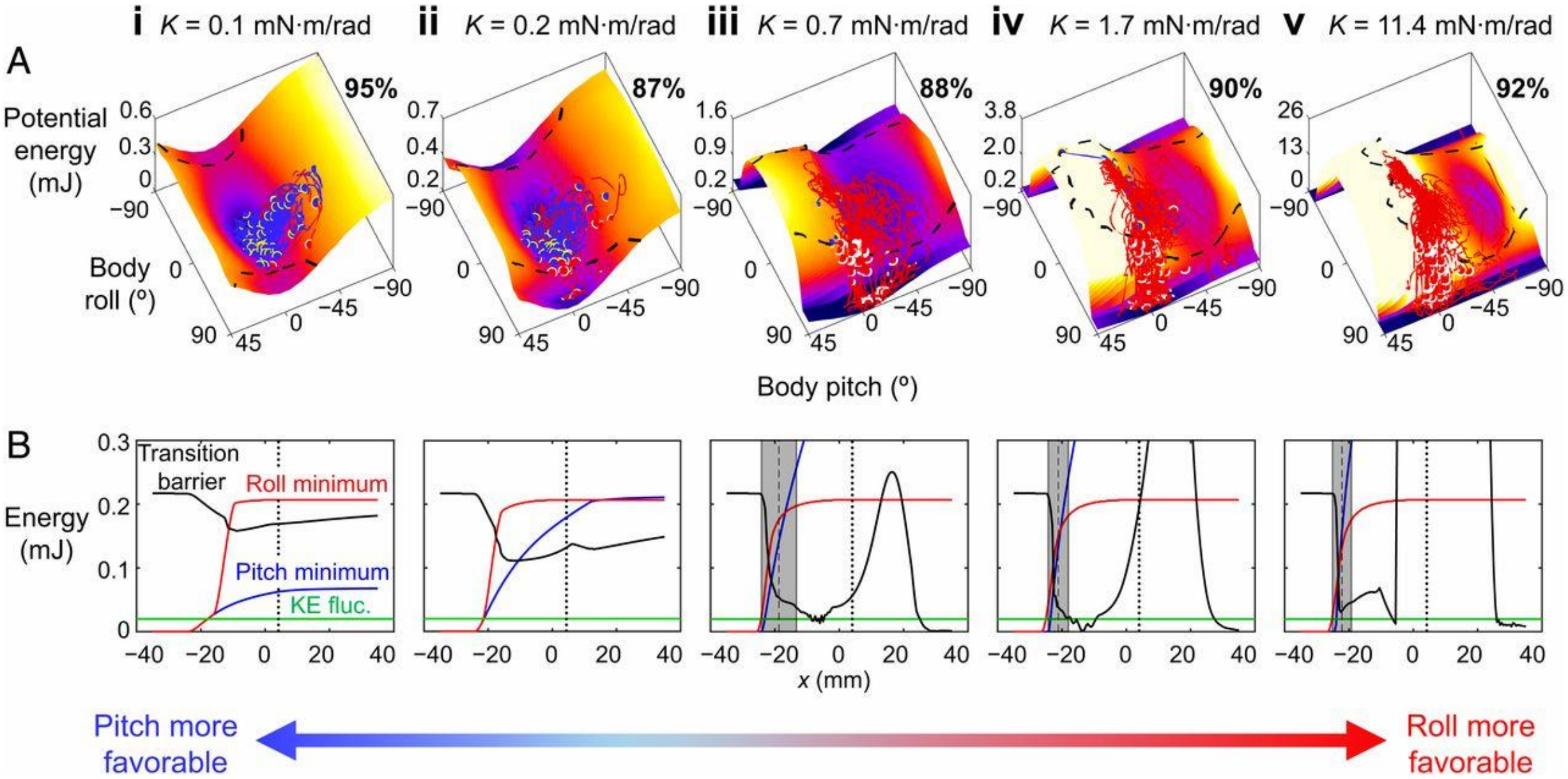

**Fig. 6.** Animal tends to transition to roll basin when kinetic energy fluctuation is comparable to potential energy barrier to escape pitch basin and when roll basin is more favorable. (*A*) Potential energy landscape over pitch-roll space (snapshot at $x = 4$ mm, dotted lines in *B*) with ensemble of state trajectories. Dashed black curves on landscape show boundary of pitch basin. Note that landscape evolves as body moves forward (increasing *x*) (SI Appendix, Movie S10), and only part of landscape over pitch-roll space is shown to focus on pitch and roll basins. We set color map scale to saturate at high energy to highlight landscape basins. (*B*) Potential energy of pitch and roll local minima and pitch-to-roll transition barrier

as a function of $x$. Green line is measured average kinetic energy fluctuation of 0.02 mJ. Columns i-v are at $K$ = 0.1, 0.2, 0.7, 1.7, and 11.4 mN·m/rad ($n$ = 64, 60, 60, 62, and 64 trials).

These results showed that physical interaction with the terrain also played a major role in the animal's probabilistic locomotor transitions, even when active behavior was likely at play. In some trials, the animal transitioned even when its average kinetic energy fluctuation was smaller than transition barrier (Fig. 6*B*). In addition, the animal occasionally transitioned to the less favorable roll mode at low $K$ (Fig. 6*A*, i, ii, red trajectories). Further, the animal often flexed its head relative to the body and used the two hind legs differentially (43) during beam interaction (23%, 63%, 89%, 79%, and 85% of the trials at the five $K$'s). All these were evidence that the animal's transition involved active behavior (see discussion). Unlike the robot that was pulled forward at a constant speed (pulling force always exceeded beam resistive force), the animal had a finite ability to push forward and may rely more on such active behavior to facilitate transition (43).

## Discussion

In summary, using a transition between two representative modes in a model system, we demonstrated that an energy landscape approach helps understand how stochastic transitions of animals and robots across locomotor modes statistically emerge from physical interaction with complex 3-D terrain. We discovered that kinetic energy fluctuation from oscillatory self-propulsion helps the system cross barriers on a potential energy landscape to make locomotor transitions. This provided compelling evidence about why variation in movement can lead to stochastic outcome (44) and can be advantageous when locomotor behavior is separated into distinct modes. This also explained early observations of surprising ability to traverse unstructured terrain of bandwidth-limited, rapid-running insects (27) and feedforward-controlled legged robots (45), as both have substantial body oscillation during locomotion. However, we view this way of "vibrate like a particle" as only one of a suite of transition strategies. Animals and robots may use other strategies to make transitions, such as plan anticipatory actions (46) and use random search (47) to overcome barriers, use sensory feedback adjustments to move towards lower barriers or reduce barriers (43), or even change morphology to modify landscape topology to introduce or eliminate certain modes (39).

We posit that there is an "energy landscape dominated" regime of locomotion, where along certain directions there exist large potential energy barriers that are comparable to or exceed kinetic energy and/or mechanical work generated by each propulsive cycle or motion. This may happen when propulsive forces are either limited by physiological, morphological, and environmental (e.g., low friction) constraints or do not well align with directions along which large barriers occur. In complex terrain with many large obstacles (34, 39, 43, 46) and even during strenuous maneuvers (35, 47–49), these situations are frequent. In this regime, not only does energy landscape modeling provide a useful statistical physics approach for understanding locomotor transitions across modes, but it may also allow comparison across systems (different animal species, robots, terrain, and modes) to discover general physical principles. Outside of this regime, energy landscape modeling is not useful—for example, not for ballistic jumping over small obstacles with kinetic energy far exceeding potential energy barriers.

We discovered that distinct attractive basins of the potential energy landscape can lead to stereotyped locomotor modes and transitions in both the animal and feedforward-controlled robot. Because our potential energy landscape is directly derived from first principles (as opposed to fitting a model to behavioral data (50–52)), this result provided compelling evidence that behavioral stereotypy of animals emerges from their neural and mechanical systems directly interacting with the physical environment (15, 16). In addition, our approach should inform how direct physical interaction with the environment constrains behavioral hierarchy (15, 16). For example, for grass-like obstacle traversal, starting with our coarse-grained landscape here resulting from a rigid body interacting with rigid "beams" on torsional springs, we can add degrees of freedom describing head flexion (43), body bending and twisting, articulated leg motions, and more realistic beam obstacles with cantilever bending and spatial heterogeneity. This will reveal more nuanced pathways of transitioning between fine-grained locomotor modes that have a variety of body and appendage configuration and terrain responses (e.g., flexing the head and tucking the legs to roll into the gap (43), separating beams laterally, etc.). Analyzing the disconnectivity (38) of basins of such a more complete, high-dimensional energy landscape will reveal the hierarchy ("treeness" (53)) of locomotor modes in complex terrain.

More broadly, these considerations suggest that our energy landscape approach provides a means towards first-principle, physical understanding of the organization of locomotor behavior, filling a critical knowledge gap. The field of movement ecology (17) makes field observations of trajectories of animals as a point mass moving and making behavioral transitions in natural environments (e.g., (54)), whose physical interactions are difficult to measure. Recent progress in quantitative ethology has advanced understanding of the organization of behavior (15, 16, 51–53), often by quantifying kinematics in homogeneous, near featureless laboratory environments (50, 51, 53, 55). Our work highlights the importance and feasibility of, and opens new avenues for, studying how the organization of behavior is constrained by an animal's direct physical interaction with realistic environments (24). Doing so will help inform how animal behavior evolves in nature; it will also simplify robot design, control, and planning to generate robust locomotor transitions in complex terrain, which may be otherwise intractable in the large locomotor-terrain parameter space. This is analogous to rugged free energy landscapes allowing divide-and-conquer in protein folding (56).

Our empirically discovered physical principles of locomotor transitions are surprisingly similar to those of microscopic systems (SI Appendix, Fig. S7), especially multi-pathway protein folding transitions where predictive energy landscape theories have been very successful (36–38). Thus, we envision our energy landscape as the beginning of a statistical physics theory that will quantitatively predict global structures and emergent dynamics of multi-pathway locomotor transitions in the energy landscape dominated regime. An immediate next step towards this is to model conservative forces using potential energy landscape gradients and add stochastic, non-conservative propulsive and dissipative forces that perturb the system to "diffuse" across landscape barriers (analogous to (57, 58)). Doing this will also elucidate how escape dynamics from a basin locally leads to emergent favorability difference between basins. These physical principles will help reveal how animals, and how robots should, use local force sensing to control motion to facilitate locomotor transitions on the landscape. Further, although it seems obvious that near-equilibrium statistical thermodynamics does not directly apply here, an energy landscape approach to locomotor transitions in complex terrain provides opportunities to test and develop new theories of few-body active matter (59).

Finally, our energy landscape approach provides a conceptual way of thinking about locomotor modes beyond near-steady-state, limit-cycle-like behavior (e.g., walk, run, climb (6–8)) by adding metastable behavior (60) locally attracted to landscape basins (e.g., pitch and roll modes here, which are far-from-steady maneuvers). We foresee the creation of new dynamical systems theories of terrestrial locomotion (25) that produce transitions across locally attractive landscape basins as well as between limit-cycle attractors (41, 61). They will enable using physical interaction to design, control, and plan basins funneled into one another to compose (62) locomotor transitions to perform high-level tasks in the real world. Terradynamics of locomotor-terrain interaction starting from first principles (24) such as illustrated here will facilitate this progress.

## Methods

### *Robot experiments*

We used a linear actuator to propel the body forward at a constant speed of 0.7 cm $s^{-1}$ and a pair of DC motors via a linkage to vertically oscillate it at a variable frequency $f$ of 0 to 6 Hz and collected a total of 280 trials. We varied $K$ of the beams by using different combinations of torsional springs in parallel.

### *Animal experiments*

We challenged the discoid cockroach to traverse a layer of beam obstacles. We tested 6 individuals and beams of the five different $K$ and collected a total of 310 trials.

### *Potential energy landscape model*

We calculated system potential energy as the sum of body and beam gravitational potential energy and beam elastic potential energy:

$$E = m_{\text{body}} g \Delta z + \frac{1}{2} m_{\text{beam}} g L (\cos\Delta\theta_1 + \cos\Delta\theta_2 - 2) + \frac{1}{2} K (\Delta\theta_1^2 + \Delta\theta_2^2) \quad (1)$$

where $m_{\text{body}}$ is body mass, $g$ is gravitational acceleration, $\Delta z$ is body center of mass height increase from its equilibrium configuration (at near zero pitch and zero roll), $m_{\text{beam}}$ is beam mass, $L$ is beam length, $K$ is beam torsional stiffness, and $\Delta\theta_1$ and $\Delta\theta_2$ are beam deflection angles from vertical. See SI Appendix, Fig. S6 for definition of variables and parameters.

See SI Appendix, Supplementary Methods for detailed methods.

## Data availability

All data are included in the main text or the supplementary information.

## Acknowledgments

Chen Li is grateful to Dan Goldman and Bob Full for early discussion that inspired this study. We are especially grateful to Reviewer #2 who nudged us to think more deeply about the broader implications of our work. We thank Qihan Xuan for many insightful discussions, comments on the manuscript, and early preliminary modeling. We thank Dan Koditschek, Bob Full, Dan Goldman, Simon Sponberg, Shai Revzen, Noah Cowan, and two anonymous reviewers for helpful suggestions; Reviewer #1 for providing torsional stiffness data of natural flexible terrain elements; Yucheng Kang, Siyuan Yu, and Jundong Yi for help with experimental setup; Yuanfeng Han and Sean Gart for help with animal care; Changxin Yan for verifying tracking accuracy; Yuanfeng Han for characterizing lens distortion; and Yaqing Wang for CAD schematic of animal locomotion arena and counting active behavior in animal trials.

This work is funded by an Army Research Office Young Investigator Program (grant # W911NF-17-1-0346), a Burroughs Wellcome Fund Career Award at the Scientific Interface, an Arnold & Mabel Beckman Foundation Beckman Young Investigator award, and The Johns Hopkins University Whiting School of Engineering start-up funds to C.L.

**Competing interests**: Authors declare no competing interests.

**Supplementary information** is available for this paper (Supplementary Methods, Figures S1-S7, Table S1 and S2, Movies S1-S10, and SI References).

Supplementary Information for

# An energy landscape approach to locomotor transitions in complex terrain

Ratan Othayoth, George Thoms, *Chen Li

*Corresponding author. Email: chen.li@jhu.edu

**This PDF file includes:**

- Supplementary Methods
- Figures S1 to S7
- Tables S1 to S2
- Legends for Movies S1 to S10
- SI References

**Other supplementary materials for this manuscript include the following:**

- Movies S1 to S10

## Supplementary Methods

### Robotic physical model

To approximate the body shape of the discoid cockroach (1), we 3-D printed an ellipsoid-like body (Fig. S1*A*, PLA plastic using UPBOX+, Tiertime, CA, USA), whose top and bottom halves were slices of an ellipsoid. The body was suspended (center of mass at 10 cm above the ground) via a custom gyroscope mechanism that allowed free body pitching and rolling (Fig. S1*B*, Movie S2). We added mass to the body so that it is bottom heavy, with body center of mass at 1.1 cm below the pitch axis and 1.6 cm below the roll axis. Body pitch and roll at static equilibrium for a freely suspended body without beam contact were near zero (pitch = 3.3° ± 0.4°, roll = 1.7° ± 0.8°; note that positive pitch is pitching downward). See Table S1 for geometric dimensions and physical properties of the body.

We used a linear actuator (Firgelli FA-HF-100-12-12, Firgelli Automation, WA, USA) to propel the body forward towards the obstacles. To introduce body kinetic energy fluctuation, we oscillated the body vertically using two DC servo motors (XM430-W350T, Dynamixel, CA, USA) via a five-bar linkage mechanism 3-D printed from PLA plastic (UPBOX+, Tiertime, CA, USA). We varied kinetic energy fluctuation by varying oscillation frequency. Our preliminary experiments showed that body oscillation along different directions did not qualitatively affect the outcome. Thus, we chose vertical oscillation to better observe response in body pitch and roll.

The body oscillated vertically along the following triangular wave trajectory (fitted from the measured $z$ position):

$$z = z_0 + Aft + N(\mu, \sigma), \qquad 0 \le t \le \frac{T}{2} \qquad \text{(S1)}$$

$$z = z_0 + A(1 - ft) + N(\mu, \sigma), \qquad \frac{T}{2} < t \le T \qquad \text{(S2)}$$

where $z$ is the vertical position of the body geometric center, $f$ is vertical oscillation frequency, $T = 1/f$ is vertical oscillation period, $A = 23.4$ mm is the average vertical position range from motor actuation, and $z_0 = 102.4$ mm is the average vertical position when there are no oscillations. We added a small noise, $N$, which is normally distributed with a mean of $\mu = 0.7$ mm and a standard deviation of $\sigma = 1.2$ mm. Kinetic energy fluctuation from this noise was small compared to that from the vertical oscillation. Its addition was to prevent the body from being stuck against the beams due to friction. The vertical oscillations induced small lateral oscillations (12% of vertical oscillation amplitude), which also helped prevent from being stuck. The motor angles were commanded using a microcontroller (Open CM 0.94, Robotis, CA, USA). We note that the animal's body oscillations are much more complex, variable, and less periodic than the robot's. It was difficult to use wave oscillations with well-defined amplitude and frequency to approximate it.

**Robot beam obstacles**

For robot experiments, we mounted two rigid beams to a fixed base (Fig. S1*A*) vertically using 3-D printed torsional spring joints (Fig. S1*C*, Movie S2). We varied $K$ by using different combinations of soft and stiff torsional springs (McMaster Carr, NJ) (Fig. S1*C*, red and cyan) in parallel. The rigid beams were laser cut from acrylic plates (VLS60, Universal Laser & McMaster-Carr, NJ, USA). We covered the beam edges using smooth plastic straw (6 mm diameter) to reduce friction between them and the body during interaction.

We characterized torsional stiffness of the stiff and soft torsional springs by measuring the restoring torque about the torsional joint as a function of joint bending angle (Fig. S1*D*) using a 3-axis force sensor (Optoforce OMD-20-FG, OnRobot, Denmark). Torsional stiffness was calculated from the slope of the linear fit (across the origin) of torque as a function of bending angle (Fig. S1*E*). By combining the stiff and soft torsional springs, we varied $K$ by over an order of magnitude ([28, 55, 255, 344] mN·m/rad). See Table S1 for geometric dimensions and physical properties of the beams.

**Robot experiment imaging**

Robot experiments were recorded using three synchronized high-speed cameras (IL5, Fastec Imaging, San Diego, CA) at 200 frames $s^{-1}$ and a resolution of 1920 × 1080 pixels. To automatically track the body and beams over the entire range of rotation, we attached BEEtags (18 mm × 18 mm) on the body (9 markers), vertical oscillation transmission (3 markers), right beam (2 markers), and left beam (5 markers). We used FasMotion software (Fastec Imaging, San Diego, CA) to save the videos to storage drives after recording for tracking and processing.

**Robot experiment protocol**

Before each trial, the body was positioned at a distance of 11 cm from the beams, and the beams were set to be vertical. We started video recording and body oscillation (for $f > 0$), waited for 1 s, and then propelled the body forward at a constant speed of 0.7 cm·$s^{-1}$ by a distance of 30 cm (maximum possible by the actuator). Body oscillation was applied (for $f > 0$) until the end of forward translation. After forward translation completed, we stopped body oscillation and video recording and moved the body to its initial position for the next trial.

At each $K$, we varied kinetic energy fluctuation by varying $f$ from 0 Hz to 6 Hz with an increment of 1 Hz. At each $K$ and each $f$, we performed 10 trials. This resulted in a total of 280 trials, with 70 trials at each $K$ across all $f$. See Table S1 for detailed sample size.

**Animals**

We chose to study the discoid cockroach, *Blaberus discoidalis*, because it dwells on the floor of tropical rainforests with dense vegetation and litter and excels at traversing complex terrain (1). We used adult male discoid cockroaches (Pinellas County Reptiles, St Petersburg, FL, USA), as females are often gravid and under different load bearing conditions. Prior to experiments, we kept the cockroaches in individual plastic containers at room temperature (24 °C) on a 12h:12h light:dark cycle. See Table S1 for dimensions and mass of the animals tested.

**Animal beam obstacles**

We custom made rigid "beams" with torsional springs at the base (Fig. S2*A*). For each beam, we sandwiched a flexible layer between two stiff layers and exposed a small portion of the flexible layer (Fig. S2*B*), which acted as torsional spring joint and allowed the beams to bend in the *x*-*z* plane. We varied the thickness of the flexible layer ([0.04, 0.05, 0.07, 0.10, 0.25] mm), to vary the torsional stiffness $K$ of the torsional joint by over two orders of magnitude ([0.1, 0.2, 0.7, 1.7, 11.4] mN·m/rad) in a similar range as natural obstacles like leaves, stalks, and grass. Polyethylene terephthalate plastic (McMaster Carr, NJ, USA) and cardstock (0.2 mm thickness, Neenah Inc., GA, USA) were used for the flexible and stiff layers and bonded using thermally bonding glue (Therm-O-Web, IL, USA) and a laminating machine (AmazonBasics, Amazon). The layer of 10 beams was laser cut (VLS60, Universal Laser Systems, AZ, USA) to have identical geometry and spacing. We characterized $K$ by measuring the restoring torque about the torsional joint as a function of beam bending angle (Fig. S2*C*) using a 6-axis force and torque sensor (Nano 43, ATI Industrial Automation, NC, USA). $K$ was calculated from the slope of the linear fit (across the origin) of torque as a function of bending angle (Fig. S2*D*). See Table S1 for geometric dimensions and physical properties of the beams. See Table S2 for comparison of torsional stiffness of our beams with natural terrain elements.

**Animal multi-camera imaging arena**

We constructed an arena for animal experiments to measure locomotor transitions (Fig. S3). Previous studies showed that animals often laterally explored beam obstacles before traversing (1). To increase experimental yield, we used 10 identical beams in an obstacle layer, which presented nine gaps (narrower than animal body width, but larger than body thickness) for the animal to traverse. All the beams were vertical without external force from the animal. The beam obstacle layer was inserted into a slit cut in the flat ground between two transparent sidewalls made of acrylic sheets. A runway funneled the animal towards the middle of beam obstacle layer to minimize the interaction with the sidewall. To facilitate traversal with minimal body yaw (on average), we arranged the beam obstacle layer to be perpendicular to the direction of animal movement; the reduced body yaw allowed us to more easily visualize how trials evolved on the potential energy landscape (see section below), which was calculated using the average body yaw from all trials. Paper cardstock covered the ground surface. We placed a dark shelter with food and water on the exit side of the obstacle layer for the animal to rest after each trial.

Animal experiments were recorded using seven synchronized high-speed cameras (N5A-100, Adimec, Netherlands) at 100 frames $s^{-1}$ and a resolution of 2592 × 2048 pixels. When interacting with the obstacles, animal body orientation varied substantially. We carefully positioned the cameras around the entire arena to cover the entire rotation range of motion, with two from back views, two side views, two isometric views, and one top view (Fig. S3*A*). We used the StreamPix software (Norpix Inc., Montreal, Canada) to automatically save the videos to storage drives as they were being recorded, after which they were converted to AVI format for tracking and processing.

To automatically track the animal and beams, we attached a 7 mm × 7 mm BEEtag (2) to the animal body and 9 mm × 9 mm BEEtags to the top and bottom ends of both sides of each beam (Fig. S3*B*). The animal BEEtag was much lighter (< 0.15 g) than the animal itself (2.6 g). It was printed onto a rounded oval cardboard to minimize interference with the obstacle traversal and attached to the dorsal surface of the abdomen using ultraviolet curing glue (Bondic, Aurora, Canada).

**Animal experiment protocol**

Before the experiment, the arena was illuminated and heated to about 43°C with six work lamps (Coleman Cable, Waukegan, IL, USA). Before each trial, the animal was placed in the starting end of the arena and allowed to settle down. We then started video recording and probed the animal with a stick with a soft tip made from paper tapes to induce it to run towards the obstacles. The animal did not always immediately traverse after running into beam obstacles. Instead, it often made multiple failed attempts to traverse and sometimes explored the obstacle layer laterally to attempt traversing at different beam gaps, before eventually traversing. Once the animal traversed and reached the shelter, we stopped video recording and allowed the animal to rest for ~10 minutes before the next trial.

We tested six animal individuals and beams of five different torsional stiffness $K$ and collected a total of 337 trials. The same six individuals were tested across all $K$. We discarded trials in which any of the following were observed: (1) the animal did not move within 10 s after it was probed; (2) the animal moved back to the starting area or did not attempt to traverse; (3) the animal used the sidewall to traverse; or (4) the animal climbed up the beams and its body and all six legs lost contact with the ground. This resulted in a total of 310 accepted trials, with approximately 10 trials for each animal at each $K$. See Table S1 for detailed sample size.

**High accuracy 3-D motion reconstruction**

To calibrate the cameras over the working space for 3-D motion reconstruction, for both robot and animal experiments, we built a calibration object with multiple markers (17 for animal and 47 for robot) using Lego bricks (The Lego Group, Denmark). We then used the direct linear transformation software DLTcal5 (3) to obtain intrinsic and extrinsic camera parameters. We used a custom MATLAB script to automatically track 2-D coordinates of the markers in each camera view using the BEEtag code (2).

Using the tracked 2-D marker coordinates from multiple camera views and camera calibration parameters, we obtained the 3-D position of the four corners of each BEEtag markers using the direct linear transformation software DLTdv5 (3), which was then used to obtain the marker frame (Fig. S3*B*). For the animal, we translated and rotated the marker frame by the measured translational ($\Delta x$ = 10 mm, $\Delta y$ = −0.2 mm, $\Delta z$ = −3 mm) and rotational (roll = 0° , pitch = 10° , yaw = 1°) offsets to obtain 3-D position and orientation of the body frame at the body geometric center, which nearly overlapped with body center of mass (4). For the robot, we used a CAD model of the body to determine the location of center of mass relative to the markers fixed to the body. Depending on which body markers were reconstructed in each video frame, we translated and rotated the reconstructed marker frame by its measured translational and rotational offsets to obtain 3-D position and orientation of the body frame at the center of mass. For both the robot and animal, we used Euler angles (yaw $\alpha$, pitch $\beta$, and roll $\gamma$, $Z-Y'-X''$ Tait-Bryan convention) to define 3-D rotation. Note that with this convention, when the body pitches upward, pitch angle is negative.

To quantify the accuracy of 3-D reconstruction using BEEtag tracking combined with Direct Linear Transformation, we 3-D printed a high-precision calibration object. The calibration object had nine BEEtag markers mounted on a horizontal plate in a 3 × 3 grid with a 7 cm grid distance, each oriented at a pitch and yaw angle of 0°, 30°, and 60°. We measured the 3-D position and orientation of each marker from 3-D reconstruction (described above) and compared them to the designed values. This demonstrated that our imaging setup achieved high accuracy in 3-D position and orientation reconstruction (s.d. of position error = 0.6 mm; s.d. of orientation error = 1.1°). We also verified that lens distortion was minimal (< 1%) using the checkboard distortion measurement method.

For each trial, we calculated body translational ($v_x$, $v_y$, $v_z$) and rotational ($\omega_\alpha$, $\omega_\beta$, $\omega_\gamma$) velocities and beam bending angles from vertical ($\Delta\theta_i$) as a function of time. Beam angle was averaged from visible tags on each beam. Considering lateral symmetry, to simplify analysis of the roll mode, we flipped all trials that rolled left to rolling right. For the animal, we offset the measured lateral positions ($y$) of each trial so that $y = 0$ in the middle of the gap that the animal traversed during the final, successful attempt.

**Definition of pitch and roll modes and pitch-to-roll transition**

We defined the robot to be in the roll mode if both beams lost contact with the body and bounced back to vertical before the distal end of the body crossed the beams ($x = 0$), and we defined it to be in the pitch mode otherwise. For the robot, body motion was highly repeatable from trial to trial, and pitch-to-roll transition always resulted in a sharp decrease in system potential energy. Thus, we defined transition to occur when system potential energy reached a peak value (Fig. S4*E*, vertical dashed line (ii)), after which it immediately reduced and eventually reached a minimal value.

We defined the animal to be in the roll mode if its body roll (absolute value) exceeded 62°, because from system geometry this was the minimal roll for the body to move through the gap between two adjacent beams without deflecting them. The animal was defined to be in pitch mode otherwise. For trials that transitioned from the pitch to roll mode, we defined transitions to occur when body roll (absolute value) exceeded 20° (Fig. S4*B*, vertical dashed line (ii)). We verified that system potential energy (Fig. S4*F*) decreased at this moment.

**Data averaging**

Because the robot was propelled forward at a constant speed, its 3-D kinematics, potential energy, and kinetic energy were a function of body forward position ($x$). To obtain average 3-D kinematics and potential energy as a function of $x$, we interpolated the measured position, orientation, and potential energy over $x$ and then averaged them across all trials at a given $K$. For the robot, we averaged lateral position $y$ and body yaw for all the trials at each $K$ for each $x$ and used this average trajectory to calculate an average potential energy landscape. For the animal, because of the high variability in body lateral position and yaw, we set both to zero when calculating the average potential energy landscape at each $x$.

Because we focused on the pitch-to-roll transition (see definition in the next section), we considered only the animal's final, successful attempt in which such a transition may occur. For the final, successful attempt, we analyzed the section of the trial starting from five frames (0.02 s) before the animal's head contacted the beams (Fig. S4, dashed vertical line (i)) until ten frames (0.05 s) after the entire body crossed the obstacle layer ($x = 0$, Fig. S4, dashed vertical line (iii)). Because the robot body was translated with a constant forward speed and always crossed the beams, for it we analyzed the section of the trial starting from when the body first contacted the beams (Fig. S4, dashed vertical line (i)) until the end of forward translation (Fig. S4, dashed vertical line (iii)).

**Kinetic energy fluctuation**

For both the robot and animal, we defined body kinetic energy fluctuation as the sum of kinetic energy due to translational and rotational velocity components other than forward motion of the body ($v_y$, $v_z$, $\omega_\alpha$, $\omega_\beta$, and $\omega_\gamma$). To calculate moment of inertia, we approximated the animal body as an ellipsoid with uniform mass distribution, considering that legs only consist less than 15% of body mass (4). For the robot, we calculated moment of inertia from a CAD model of the body with accurate geometry and mass distribution.

For both the robot and animal, we calculated average kinetic energy fluctuation from first contact (see definition below) to when transition occurred using the trials that transitioned to the roll mode. This was because for the trials that was stuck in the pitch mode, it was difficult to define the onset of pitching as can be readily done for the onset of rolling. Including these trials would add the substantial kinetic energy of continuous body pitching that resulted from the interaction, which was not part of the fluctuation that induced the transition. We verified that kinetic energy fluctuation differed little between before contact with the beams and from first contact to when transition occurred. We then averaged kinetic energy fluctuation over time for each trial, from when the body first contacted a beam (Fig. S4, dashed line (i)) (for the animal, the first contact at the beginning of the final successful attempt of each trial), to when it transitioned to roll mode (Fig. S4, dashed line (ii)). For the robot, we then averaged these trial averages across all trials at each $f$ in which the robot transitioned to the roll mode to obtain average kinetic energy fluctuation at each $f$ (Fig, S5*A*). For the animal, we averaged these trial averages across all trials at each $K$ to obtain average kinetic energy fluctuation at each $K$ (Fig. S5*B*).

**Statistics**

All probabilities were calculated relative to the total number of accepted trials of each treatment. All other average data are reported as mean ± 1 s.d. For both the robot and animal, we used a mixed-effects chi-square test to test whether pitch-to-roll transition probability depended on $K$. For the robot, we included $K$ and $f$ as fixed effects; for the animal, we included $K$ and individual as fixed effects, and also considered their crossed effect. For the robot, we used an ANOVA to test whether kinetic energy fluctuation increased with $f$. For the animal's kinetic energy fluctuation, we pooled data from all the trials in which pitch-to-roll transition occurred (see section above for explanation) for each $K$ and included individual as a random factor to account for individual variability (i.e., a mixed-effects ANOVA). We used a Student's $t$-test to test whether the percentage of robot trajectories attracted to the corresponding mode basin was significantly different from 1. All statistical tests were performed using JMP Pro 13 (SAS Institute Inc., NC, USA).

**Potential energy landscape**

In energy landscape modeling, we approximated the animal body as a rigid ellipsoid and obtained the robot body shape from a CAD model used for 3-D printing the body. The beams were modeled as rigid rectangular plates on torsional joints (Fig. S6*A*). Because the beams had a finite mass, forward deflection lowered beam center of mass and thus beam gravitational potential energy. Because the measured beam restoring torque was nearly proportional to bending angle for both the robot (Fig. S1*D*) and animal (Fig. S2*C*), we approximated the torsional joint at the base of each beam as a perfect Hookean torsional spring and assumed there was no damping. Because the body only pushed forward against the beams, in the model we only allowed forward beam deflection ($\Delta\theta_{1,2} \geq 0$).

For the robot, we set center of mass to be below the pitch and roll axes (Fig. S6*B*). For animal modeling, we constrained the lowest point of the body to always touch the ground (ground constraint) (Fig. S6*C*), because the animal maintained ground contact during traversal (we rejected trials in which the animal climbed onto the beams). The potential energy calculated from the model was only an approximation of the actual potential energy because we neglected the animal's legs. Thus, for both the robot and animal, body pitching and rolling in response to interaction with the beams would increase center of mass height and thus body gravitational potential energy. In addition, because the robot was suspended from and driven forward by a linear actuator, its center of mass height was constrained to move within a measured range of $z$ = [9.9 cm, 11.8 cm]. Considering that the robot's controlled vertical oscillation was modeled as part of kinetic energy fluctuation, we used the average body center of mass vertical position before contacting the beams ($z$ = 10.8 cm) to calculate its initial body potential energy; we verified that at any given $x$, landscape shape remained similar within the $z$ range in which the robot was oscillated.

For both the robot and animal, we offset system potential energy to zero when the body was not in contact beams and in its static equilibrium (at zero pitch and zero roll), and system potential energy shown on the landscapes were relative to this initial equilibrium (Fig. S6*B*, *C*).

The full potential energy landscape depended on body orientation (pitch, roll, yaw) and forward and lateral positions ($x$, $y$), given the vertical height and ground constraints on the robot and animal, respectively. Because we focused on body pitch and roll motions, for a given body position ($x$, $y$) and yaw, we varied body pitch and roll over [−180°, 180°] to calculate system potential energy landscape over pitch-roll space. In Figs. 2*B*, 3, 4*A*, and 6*A*, we only show the landscape over part of the entire pitch-roll space to better focus on the pitch and roll basins. We then calculated beam deflection due to body contact (only allowing $\Delta\theta_{1,2} \geq 0$) and center of mass height increase ($\Delta z$) to obtain system potential energy using Eqn. 1 (main text). We note that our landscape did not model body-beam interaction after the beams bounced back.

### Local minima and system state trajectories on potential energy landscape

For each forward position of the body relative to the beams ($x$), we examined the landscape to determine the pitch and roll local minima and measured their potential energies. Note that for the robot their potential energies did not include height change due to controlled vertical oscillation (see section above). To visualize how the measured state of the system behaved on the landscape, we projected the measured body pitch and roll onto the landscape for each $x$ (Figs. 3*A*, 4, Movies S4, S6, S8, blue and red dots for pitch and roll modes), which formed a system state trajectory over time as traversal progressed. Note that only the end points of the trajectory, which represent the current state, showed the actual potential energy of the system at the corresponding $x$. The rest of the visualized trajectory showed how body pitch and roll evolved, but for visualization purpose was simply projected on the landscape surface. Because roll local minimum does not exist at $K$ = 28 mN·m/rad, for comparison with other $K$, we defined it to be at (pitch, roll) = (0°, ± 42°) based on the minimal body roll required to traverse without beam bending.

**Average potential energy landscape at each beam stiffness**

To facilitate observation of statistical trends, we calculated an average potential energy landscape at each $K$ and visualized all trials on it. Average landscape calculation used the average measured lateral position ($y$) and body yaw for each $x$. For the robot, this average potential energy landscape was a good approximation of the actual landscape for each trial, because the robot was constrained by design to have minimal lateral motion or yawing. Despite this, when projected onto the average potential energy landscape, in some trials at high $K$, a portion of the system state trajectory appeared to momentarily go out of the pitch basin and then re-entered it (Figs. 4, 6*A*, Movies S5, S6, S8, S10). This was an artifact from landscape averaging. In those trials, the body experienced larger yawing due to a slight lateral bending of the plastic pole that suspended the robot resulting from high beam restoring forces. Because such trials are rare, the average landscape basin was close to that without body yawing. Examination of the actual landscape for each trial (see next section) verified that the state trajectory in the pitch mode was almost always in the pitch basin. For the animal that freely moved laterally and yawed, the average landscape was a much poorer approximation of the actual landscape for each trial.

**Percentage of trials attracted to basin of observed mode on actual landscape**

Because the average landscape did not account for trial-to-trial variation, to better quantify how well the potential energy landscape explained the observed locomotor modes, for both the robot and animal, we further reconstructed the actual (not averaged) potential energy landscape for each trial using the measured position ($x$, $y$) and body yaw of that trial (see Movie S6 for examples). We then counted the number of trials in which the system state either stayed in the pitch basin or transitioned to the roll basin, in accord with the locomotor observed, and we calculated the percentage of trajectories attracted to the corresponding basin.

**Energy barrier to escape pitch minimum basin**

We quantified the potential energy barrier that must be overcome to escape from the pitch local minimum. First, at each body forward position ($x$), we considered imaginary straight paths away from the pitch local minimum (Fig. 2*B*, iii, blue dot) in the full pitch-roll space ([−180°, 180°]), parameterized by an angle $\psi$ relative to the negative pitch direction (body pitched up). Along each imaginary straight path, we obtained a cross section of the potential energy landscape (Fig. 2*B*, iii, inset). Then, we measured and defined the maximal increase in potential energy in the cross section as the escape barrier along this imaginary straight path, which was a function of $\psi$, as shown by a polar plot (Fig. 4*B*, Movie S7, top row). Then, we calculated how escape barrier vs. $\psi$ changed as traversal progressed (increasing $x$) (Movie S7, bottom row). We defined the lowest escape barrier as the pitch-to-roll transition barrier (Movie S7, top right panel, black circular arc) and measured the roll and pitch angle of its location as $x$ increased. For the robot, we calculated pitch-to-roll transition barrier using the average landscape at each $K$. For the animal, we used an average landscape with zero average lateral position and body yaw, considering its large trial-to-trial variation in lateral position and body yaw.

**System state velocity directions**

To measure the direction towards which the robot state trajectory was moving in the pitch-roll space during transition, for each trial, we calculated the velocity vector of the state trajectory in the pitch-roll space from the measured body roll and pitch and low-pass filtered the data using a sixth order Butterworth filter (Movie S9, bottom). Then, we calculated the polar angle of this velocity vector relative to the pitch-roll axes of the landscape. To focus on the transition, for trials in which pitch-to-roll transition occurred, we only considered the portion of the trial occurring over the $x$ range from start of beam contact to the onset of transition (Fig. S4, vertical dashed lines i-ii); for trials in which the transition did not occur, we considered the portion of the trial within the average $x$ range where transition was observed at higher $K$ ($x = [-69, -39]$ mm). For each $K$, we pooled data of trials in which the system was stuck in the pitch mode and those in which the system transitioned to the roll mode to calculate their respective distribution (polar histogram) of velocity directions (Fig. 4*D*). We also measured the directions of the saddle point between the pitch and roll basins and the local maximum along the pitch-up and pitch-down directions, averaged over the $x$ range in which transition was observed (Fig. 4*D*, brown and gray dashed lines).

**Active body and limb adjustments**

We observed high speed videos of animal experiments to search for evidence of the animal using active adjustments to make transition (5). For each $K$, we counted the percentage of trials in which the animal repeatedly flexed its head, differentially used its hind legs, or did both.

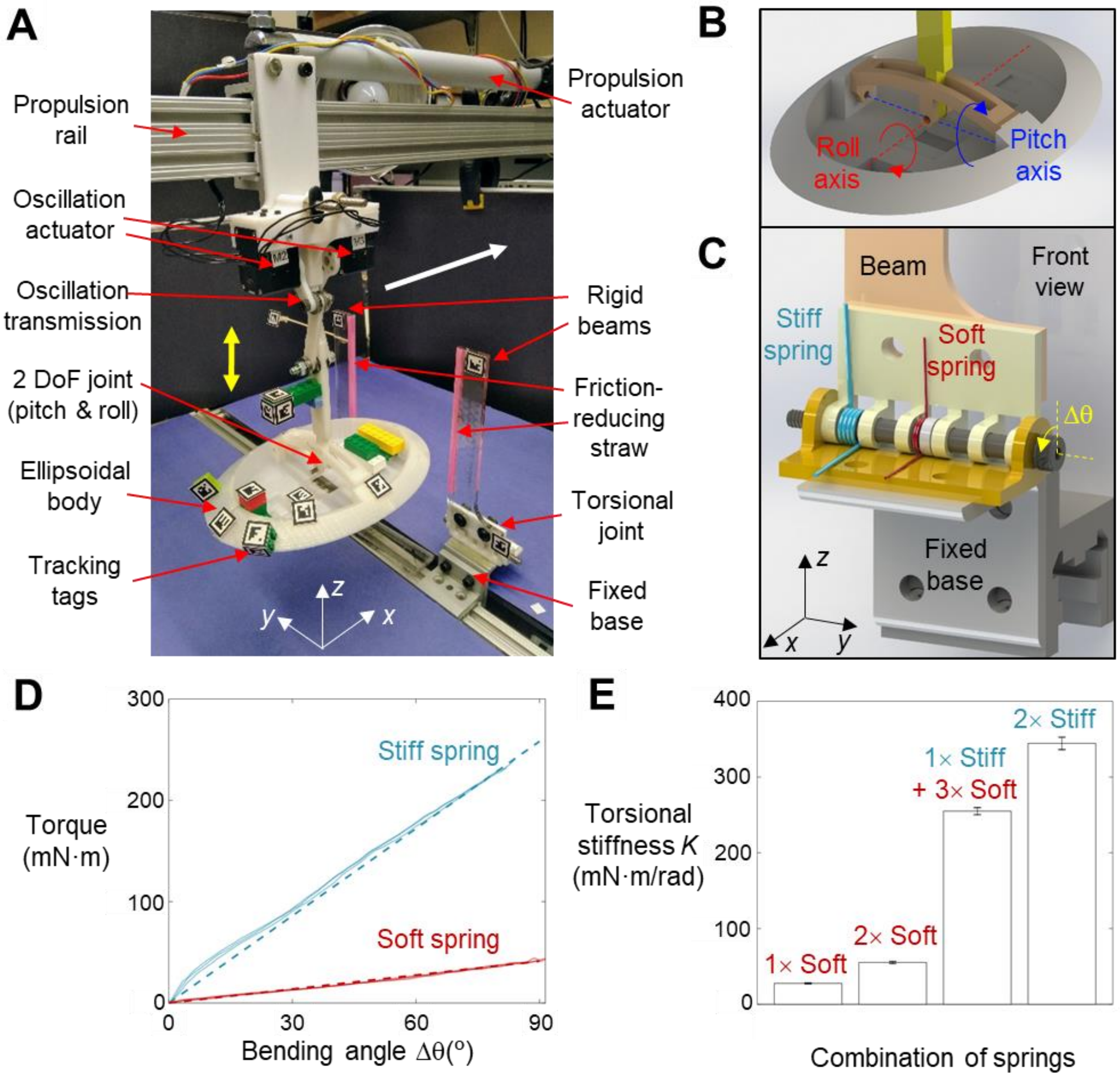

**Fig. S1.** Design of robotic physical model and rigid beams with torsional springs at base and robot beam stiffness characterization. (*A*) Photo of robot body and beams. Body is propelled forward at a constant speed (white arrow) and can be oscillated vertically (yellow arrows). Body can freely pitch and roll in response to interaction with beams. (*B*) CAD model of body, showing design of pitch and roll joints and axes. Body center of mass is below geometric center due to added weight. Pitch (blue) and roll (red) axes cross geometric center. (*C*) CAD model of beam base, showing design of torsional joint. Rigid beams rotate about an axis parallel to *y*-axis (yellow arrow). *K* is varied by using different combinations of soft (red) and stiff (cyan) springs. (*D*) Beam restoring torque as a function of bending angle Δθ (defined in *C*). Red and cyan curves are data for soft and stiff spring. Dashed lines are linear fits (through the origin) of data of each *K*,

whose slope give $K$. (*E*) $K$ for different combinations of springs used (mean ± s.d., $n = 3$ springs, 3 loading cycles each).

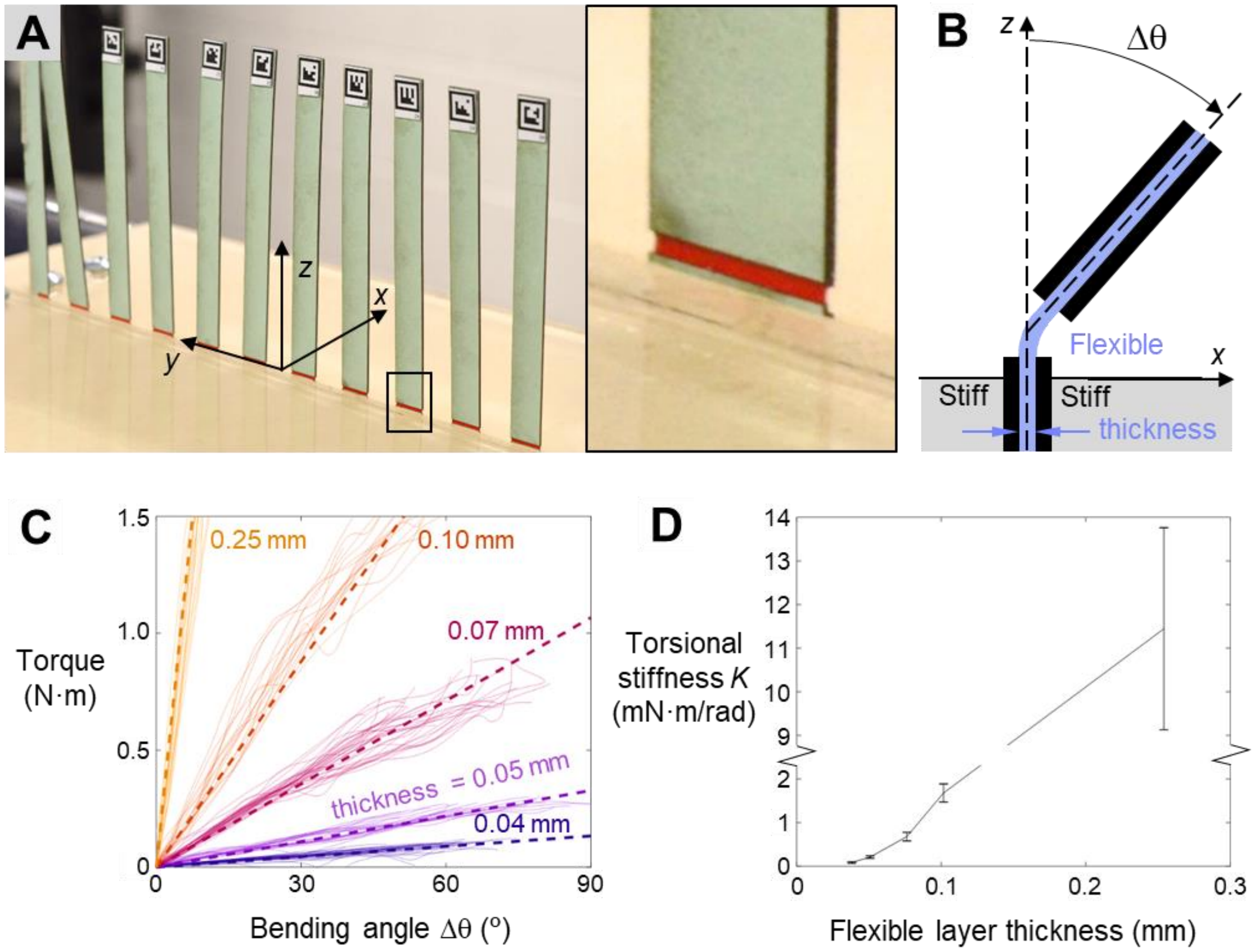

**Fig. S2.** Design and characterization of rigid beams with torsional springs at base for animal experiments. (*A*) Photo of a layer of animal beams. Inset shows a closer view of torsional joint. (*B*) Side view schematic of beam design following (6). Stiff outer layers (black) provide rigidity, and a small exposed section of flexible inner layer (blue) acts as a torsional spring joint. Dimensions not true to scale. (*C*) Beam restoring torque as a function of bending angle for different flexible layer thickness ([0.04, 0.05 0.07, 0.10, 0.25] mm). Dashed lines are linear fits (through the origin) of data, whose slopes give *K*. (*D*) *K* as a function of flexible layer thickness (mean ± s.d., $n$ = 62, 37, 76, 38, 32 loading cycles).

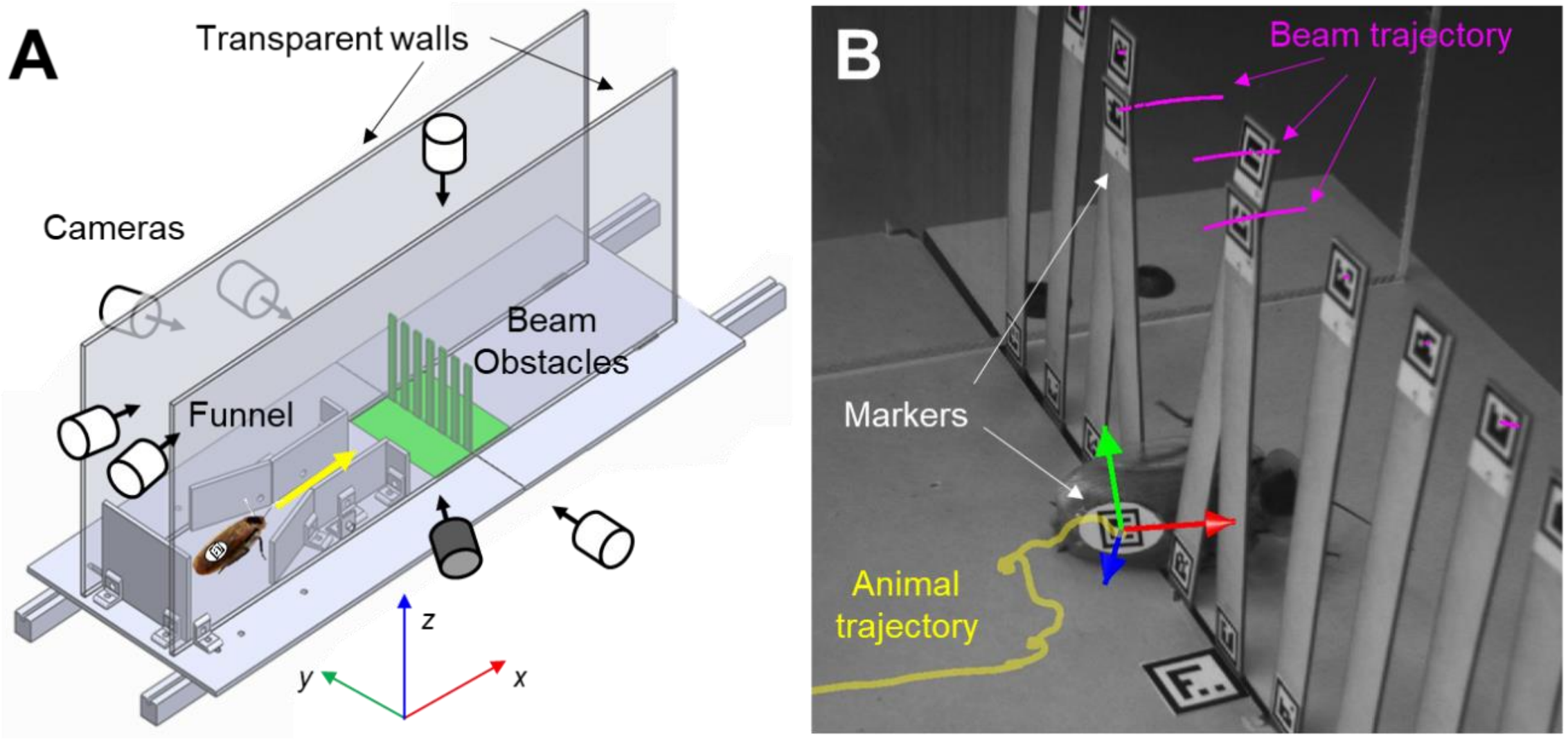

**Fig. S3.** Animal experimental setup. (*A*) Animal locomotion arena with a layer of beam obstacles (green), with seven high-speed cameras. *x*, *y*, *z* axes show lab frame. (*B*) Snapshot of animal traversing beam obstacles (view from the shaded camera in *A*). Markers are attached to the animal body and beams to track their 3-D motion (yellow and magenta trajectories). Red, green, blue axes show body frame attached to markers.

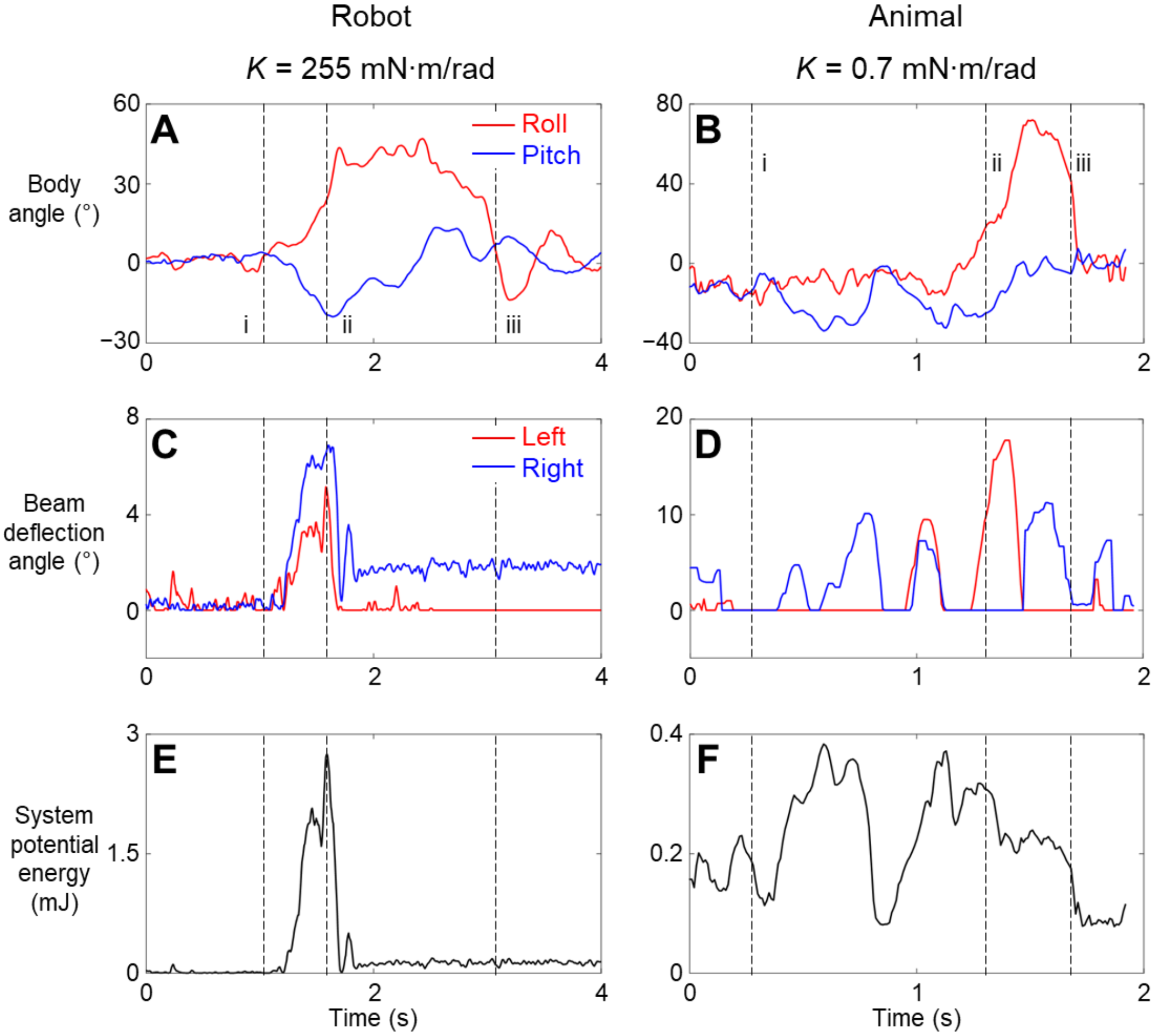

**Fig. S4.** Representative motion of body and beams and system potential energy during interaction and definition of traversal and pitch-to-roll transition. (*A*, *B*) Body roll (red) and pitch (blue) as a function of time. (*C*, *D*) Left (red) and right (blue) beam deflection angle as a function of time. (*E*, *F*) System potential energy as a function of time. Data shown for a representative pitch-to-roll transition at $K$ = 255 mN·m/rad for robot and $K$ = 0.7 mN·m/rad for animal. For both the robot and animal, pitch-to-roll transition resulted in a reduction in system potential energy. Note that negative pitch is the body pitching head-up. Dashed lines (i) and (ii) are when body first contacts beams and when pitch-to-roll transition occurs. Dashed line (iii) is when the robot's forward translation ends and when animal's distal end crosses the beam ($x$ = 0).

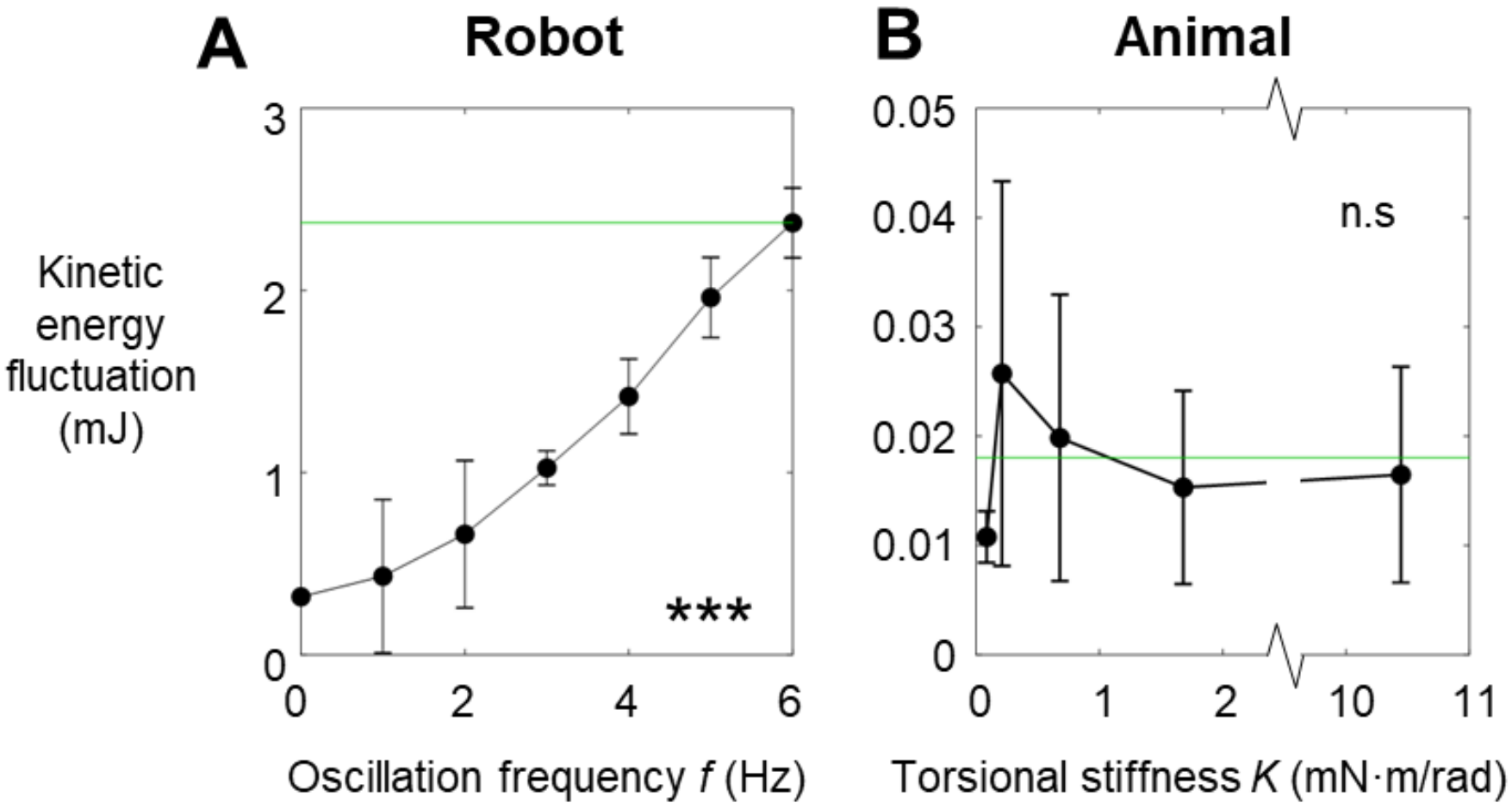

**Fig. S5.** Kinetic energy fluctuation. (*A*) Kinetic energy fluctuation of robot as function of *f*. *** indicates a significant dependence (ANOVA, $P$ < 0.0001, $F$ = 520.99). (*B*) Kinetic energy fluctuation of animal as function of *K*. n.s. indicate no significant difference (ANOVA, $P$ = 0.3835, $F$ = 0.9047). See Table S1 for sample size.

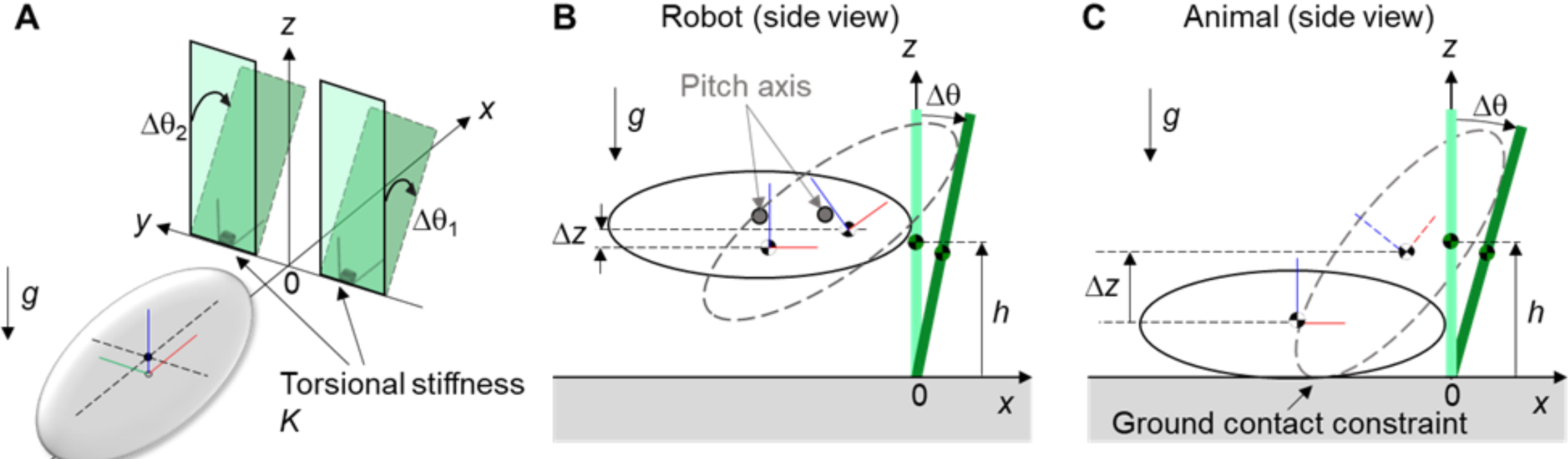

**Fig. S6.** Potential energy landscape model, with definition of variables and parameters. (*A*) Oblique view schematic of body (a rigid ellipsoid) and beams (rigid rectangular plates with torsional joints at base) of torsional stiffness *K*. Without body contact, both beams are vertical (light green). With body contact, beams are deflected forward (dark green) by angles $\Delta\theta_{1,2}$. (*B*, *C*) Side view of model for robot (*B*) and animal (*C*) to show center of mass height changes with body pitching and beam deflection. Solid and dashed ellipses show body in static equilibrium and pitched-up, respectively. Center of mass of body and beams are shown.

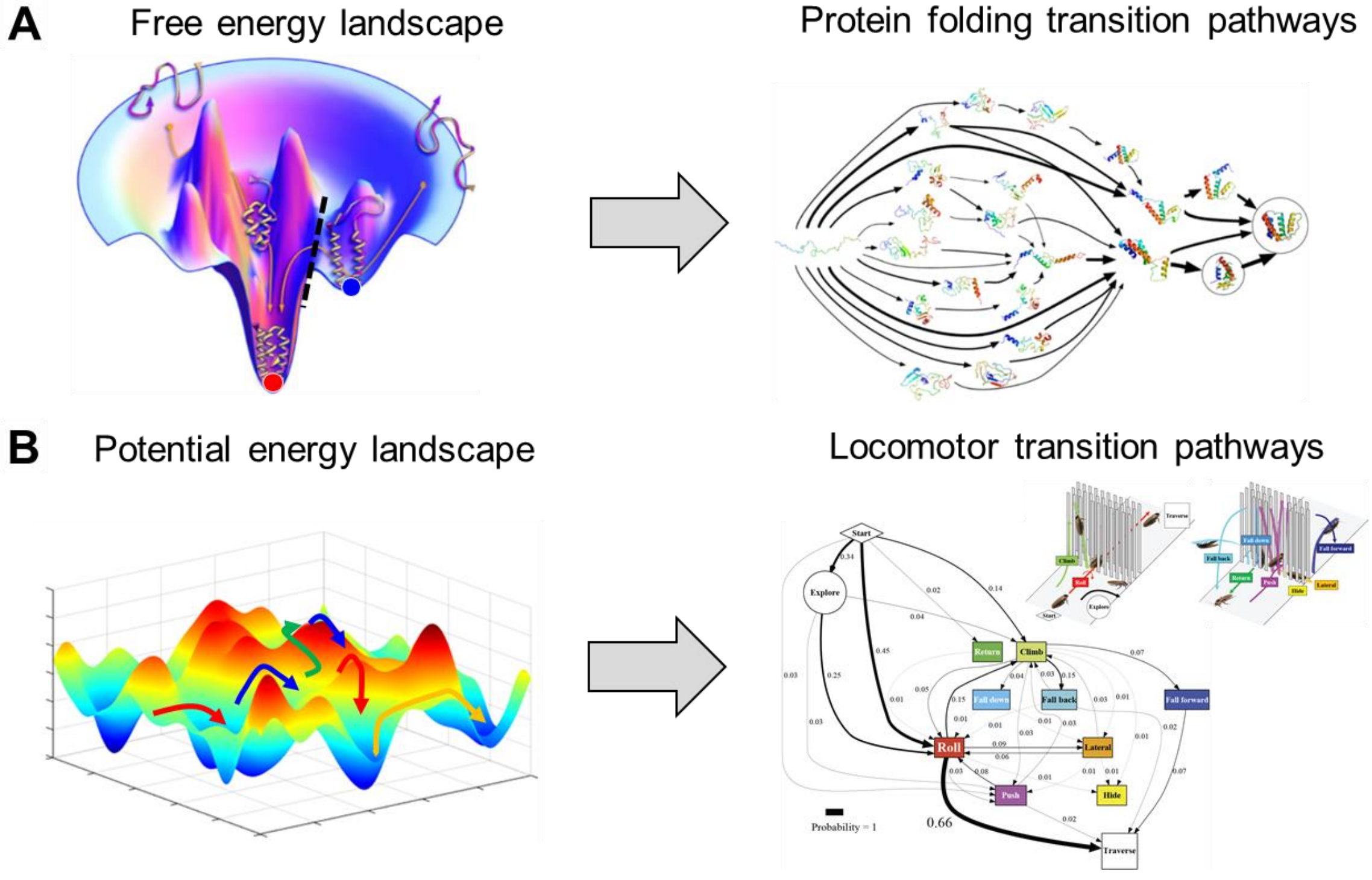

**Fig. S7.** Comparison of energy landscape between protein-folding transitions and locomotor transition. (*A*) Energy landscape theories help understand physical principles and predict global structures and emergent properties of probabilistic protein folding transitions via multiple pathways (7–10). Image credits: *A*, left panel from (10). Reprinted with permission from AAAS; *A*, right panel: Adapted with permission from (11). Copyright (2012) American Chemical Society. (*B*) We envision energy landscape modeling as a beginning of a statistical physics approach for understanding and predicting probabilistic, multi-pathway locomotor transitions in complex terrain (1).

**Table S1. Geometric dimensions, physical properties, and sample size for animal and robot experiments**.

| | | **Animal** | | | | | | **Robot** | | | | |
|---|---|---|---|---|---|---|---|---|---|---|---|---|
| | Number of individuals | 6 | | | | | | N/A | | | | |
| **Body** | Mass $m_{body}$(g) | 2.6 ± 0.3 | | | | | | 233 | | | | |
| | Length (cm) | 5.3 ± 0.1 | | | | | | 22.1 | | | | |
| | Width (cm) | 2.4 ± 0.1 | | | | | | 15.8 | | | | |
| | Thickness (cm) | 0.8 ± 0.1 | | | | | | 5.8 | | | | |
| **Beam** | Lateral spacing (cm) | 1 | | | | | | 12.7 | | | | |
| | Width (cm) | 1 | | | | | | 2.8 | | | | |
| | Mass $m_{beam}$ (g) | | 0.33 | 0.42 | 0.63 | 0.70 | 1.03 | | 38 | | | |
| | Inner layer thickness (mm) | | 0.04 | 0.05 | 0.07 | 0.10 | 0.25 | | N/A | | | |
| | Total thickness (mm) | | 0.54 | 0.55 | 0.72 | 0.75 | 0.85 | | 6 | | | |
| | Height $2h$ (cm) | | 5.7 | 8.8 | 8.7 | 8.6 | 9.3 | | 18 | | | |
| | Torsional stiffness $K$ (mN·m/rad) | | 0.1 | 0.2 | 0.7 | 1.7 | 11.4 | | 28 | 55 | 255 | 344 |
| **Sample size** | Number of trials | Ind. 1 | 11 | 10 | 9 | 11 | 10 | 0 Hz | 10 | 10 | 10 | 10 |
| | | Ind. 2 | 10 | 10 | 10 | 7 | 10 | 1 Hz | 10 | 10 | 10 | 10 |
| | | Ind. 3 | 11 | 10 | 10 | 11 | 11 | 2 Hz | 10 | 10 | 10 | 10 |
| | | Ind. 4 | 13 | 10 | 11 | 11 | 13 | 3 Hz | 10 | 10 | 10 | 10 |
| | | Ind. 5 | 10 | 10 | 10 | 12 | 10 | 4 Hz | 10 | 10 | 10 | 10 |
| | | Ind. 6 | 9 | 10 | 10 | 10 | 10 | 5 Hz | 10 | 10 | 10 | 10 |
| | | Total | 64 | 60 | 60 | 62 | 64 | 6 Hz | 10 | 10 | 10 | 10 |
| | Total number of trials | 310 | | | | | | 280 | | | | |

All data averages are mean ± 1 s.d. unless otherwise specified.

**Table S2. Torsional stiffness of our beams in comparison with natural obstacles**.

| **Source** | **Plant / Entity** | **Torsional Stiffness (mN·m/rad)** |
|---|---|---|
| Niklas, 1991 (12) | Trembling poplar | 0.1 |
| *This study* | *K1* | 0.1 |
| *This study* | *K2* | 0.2 |
| Vogel, 1992 (13) | Sunflower | 0.7 |
| *This study* | *K3* | 0.7 |
| Vogel, 1992 (13) | White poplar | 1.2 |
| *This study* | *K4* | 1.4 |
| Vogel, 1992 (13) | Red maple | 1.9 |
| Vogel, 1992 (13) | Green bean | 6.8 |
| Vogel, 1992 (13) | Sweet gum | 9.8 |
| *This study* | *K5* | 11.4 |
| Gibson et al, 1988 (14) | Iris leaf | 21 |
| O'Dogherty and Gale, 1991 (15) | Grass | 73 |
| Ennos, 1993 (16) | Sedge | 77 |
| Vogel, 1992 (13) | Cucumber | 87 |
| Vogel, 1992 (13) | Tomato | 120 |
| Etnier, 2001 (17) | Horsetail | 187 |
| O'Dogherty and Gale, 1991 (15) | Straw | 347 |

To assess how our beams compared with natural flexible terrain elements, we surveyed literature of a diversity of natural terrain elements. We estimated the torsional stiffness of each terrain element by considering a cantilever beam made from it, using $K = \frac{EI}{L}$, where $E$ is the Young's modulus, $I = \frac{1}{12} L \times d^3$ is the moment of inertia of the cross section of the beam, $L$ is the lateral width of the beam, and $d$ is the thickness of the beam. We obtained $E$, $I$, $L$, and $d$ from studies. Where $L$ is not available, we set it to be L = 1 cm, the same as our beams. We found that our beam obstacles are representative. In addition, because our energy landscape model is from first principles, the principles are likely general (beyond the stiffness range tested).

Movie S1. **Locomotor transitions of a cockroach in complex terrain.** Modes shown are only for illustrative purpose.

Movie S2. **Robotic physical model.** The body is free to pitch and roll and the beams can deflect about joints with torsional springs at the base.

Movie S3. **Animal traversing beam obstacles in pitch mode or transitioning to roll mode.** Yellow and pink curves are trajectories of animal body and top end of beams, respectively. Red, green, and blue arrows define animal's body frame. Inset is reconstructed 3-D motion.

Movie S4. **Robot traversing beam obstacles in pitch mode or transitioning to roll mode.** Left inset: reconstructed 3-D motion. Right: reconstructed potential energy landscape with current state (dot) and state trajectory.

Movie S5. **Robot state trajectory ensemble on average potential energy landscape.** Top inset is 3-D reconstruction of a representative trial without oscillation ($f = 0$). Trials rolling left are flipped to rolling right considering lateral symmetry.

Movie S6. **Robot state trajectory on actual potential energy landscape of representative trials.** Each row shows five trials at a different beam stiffness. Trials rolling left are flipped to rolling right considering lateral symmetry.

Movie S7. **Robot potential energy barrier to escape from pitch local minimum.** Top row shows landscape, landscape cross section, and escape barrier from pitch local minimum along different directions in pitch-roll space, at $x = -46.5$ mm. Bottom row shows evolution of landscape, pitch-to-roll transition barrier (lowest escape barrier), and escape barrier along different directions, as body moves forward (increasing $x$). Left: potential energy landscape. Blue and red dots are pitch and roll local minima. Green

circle on right panels and green line in bottom middle panel show average kinetic energy fluctuation of 2.3 mJ at $f = 6$ Hz. Yellow dot in left and right panels show saddle point between pitch and roll basins.

**Movie S8. Comparison of robot pitch-to-roll transition on potential energy landscape across beam stiffness.** Top: Robot state trajectory ensemble on potential energy landscape. Middle: Escape barriers from pitch local minimum along different directions. Bottom: Pitch-to-roll transition barrier (lowest escape barrier). Gray band shows $x$ range in which pitch-to-roll transition is observed (mean ± s.d.). Green circle in middle panels and green line in bottom panels show average kinetic energy fluctuation of 2.3 mJ at $f = 6$ Hz. Trials rolling left are flipped to rolling right considering lateral symmetry.

**Movie S9. Robot escape from pitch basin is more likely towards direction of lower barrier.** Top: System state trajectories in pitch-roll space. Bottom: System state velocity directions in pitch-roll space. Green and yellow dots show pitch local minimum and saddle point between pitch and roll basins. Black curve is boundary of pitch basin.

**Movie S10. Animal pitch-to-roll transition on potential energy landscape.** Top: Representative video. Middle: Animal state trajectory ensemble on potential energy landscape. Bottom: Pitch-to-roll transition barrier (lowest escape barrier). Gray band shows $x$ range in which pitch-to-roll transition is observed (mean ± s.d.). Green line in bottom panels show average kinetic energy fluctuation of 0.02 mJ. Trials rolling left are flipped to rolling right considering lateral symmetry.

**SI References**